\documentclass[%
reprint,
superscriptaddress,
amsmath,amssymb,
aps,
pra,
]{revtex4-2}
\usepackage{bm}
\usepackage{graphicx}
\usepackage[hidelinks]{hyperref}
\usepackage{comment,xcolor}
\usepackage{mathrsfs}
\usepackage[utf8]{inputenc}
\usepackage{array,leftindex}

\newcolumntype{P}[1]{>{\centering\arraybackslash}p{#1}}

\DeclareMathOperator{\sech}{sech}

\begin{document}

\def\be{\begin{equation}}
\def\en#1{\label{#1}\end{equation}}
\newcommand{\rd}{\mathrm{d}}
\newcommand{\vare}{\nu }
\newcommand{\tb}{\mathbf{t}}
\newcommand{\Phib}{\mathbf{\Phi}}
\newcommand{\Psib}{\mathbf{\Psi}}
\newcommand{\U}{\mathcal{U}}

\title{Isoenergetic model for optical downconversion and error-specific limits of the parametric approximation}
\author{D. B. Horoshko}
\affiliation{Institut f\"ur Quantenoptik, Universit\"at Ulm, Ulm D-89081, Germany}

\author{V. S.  Shchesnovich}
\affiliation{Centro de Ci\^encias Naturais e Humanas, Universidade Federal do
ABC, Santo Andr\'e,  SP, 09210-170 Brazil}

\begin{abstract}
 
Optical downconversion is widely used for generating photon pairs, squeezed and entangled states of light, making it an indispensable tool in quantum optics and quantum information. In the regime where the pump is much stronger than the generated field, the standard parametric approximation treats the pump amplitude as a fixed parameter of the model. This approximation has a limited domain of validity since it assumes a non-depleted and non-entangled pump. By finding an approximate solution to the Schrödinger equation of the downconversion process, we obtain an improved analytical model beyond the parametric one, which accounts for pump depletion and pump-signal entanglement. The new model is advantageous, first, because it allows one to compute averages of field operators far beyond the domain of validity of the parametric approximation, and second, because it allows one to establish error-specific limits of the latter domain. For a given pump amplitude, we find a maximum squeezing parameter, up to which the approximation remains valid within a specified acceptable error. Our results confirm that recent experiments on Gaussian boson sampling, with a squeezing parameter of $r\approx  1.8$ and a coherent pump amplitude of $\alpha\approx 2\cdot10^6$, can still be accurately described by the standard parametric approximation. However, we observe a sharp decline in validity as the squeezing parameter increases. For pump amplitudes of $\alpha \approx 2\cdot10^6$, the parametric approximation breaks down when the squeezing parameter exceeds $r\approx  4.5$, whereas the new approximation remains valid up to $r\approx 6$ with an acceptable error of 1\%.
 \end{abstract}

\maketitle

\section{Introduction}  

Optical parametric downconversion \cite{KlyshkoBook,Mandel&Wolf} is a versatile tool for classical optical applications such as frequency conversion and parametric amplification, as well as an essential source of entangled and squeezed states of light for photonic quantum technologies \cite{OBrien09}. In most applications, the pump is treated as an undepleted classical wave that modulates the coupling parameter of the subharmonic fields, an approach known as the parametric approximation \cite{Louisell61}. This approximation is sufficient in the low-gain regime, where single photon pairs are generated \cite{Hong85,Fabre20}, and in the high-gain regime with a moderate degree of squeezing \cite{Yuen76,Loudon87,Weedbrook12}. However, many quantum information tasks, such as quantum-enhanced interferometry \cite{Schnabel17,Hashimoto24}, cluster state quantum computation \cite{Larsen19,Asavanant19}, and demonstrations of quantum advantage through Gaussian boson sampling \cite{Aaronson11,Lund14,Hamilton17}, require a high degree of squeezing. Progress in this area is very fast, and current experiments approach the quantum advantage regime \cite{Zhong20,Zhong21,Deng23,Madsen22}. Although higher squeezing levels can be achieved by increasing the pump power and using longer nonlinear crystals, this also results in a larger fraction of the pump energy being transferred to the signal field, thereby challenging the validity of the parametric approximation.

The limits of the parametric approximation were first established in the seminal work of Hillery and Zubairy \cite{Hillery84}, who analyzed generation of a single-mode signal field under the interaction Hamiltonian
\be
{H}= i\hbar\kappa \left( {a} {b}^{\dag 2}- {a}^\dag {b}^2\right),
\en{Ham}
where $a$ and $b$ are the photon annihilation operators for the pump and signal modes, respectively, and $\kappa$ is the coupling constant proportional to the nonlinear susceptibility of the crystal. In the parametric approximation, the pump is assumed to be in a coherent state with amplitude $\alpha$. This allows one to make the substitution $a\to\alpha e^{-i\omega_p t}$, where $\omega_p$ is the pump frequency, leading to a parametric Hamiltonian that transforms the initial vacuum state of the signal mode into the squeezed state $|r\rangle$ with the squeezing parameter $r$.

Using a path-integral approach, Hillery and Zubairy computed certain correlation functions for the signal-mode operators up to the first order in a perturbative expansion, where the zeroth-order term corresponded to the parametric approximation. They also identified the conditions under which this approximation remains valid: specifically, when each component of the vector $(1/\alpha, r/2\alpha, re^{2r}/2\alpha, e^{2r}/\alpha)$ is much smaller than 1. As noted by the authors, these conditions are somewhat redundant and can be rewritten more concisely as \mbox{$\max(2,r)e^{2r}/2\alpha\ll1$.} While this limitation has been sufficient for quantum optical experiments over the past 40 years, the increasing demands of modern applications now call for a more refined limitation.

Many modern quantum technologies require approximation limits to be error-specific. For example, a Gaussian boson sampler \cite{Lund14,Hamilton17} enables observation of the statistical frequency $F_k$ of the $k$th outcome in the distribution of a certain number of indistinguishable bosons across a given set of modes. This frequency asymptotically approaches the probability $P_k$ predicted by the quantum model. For sufficiently large numbers of bosons and modes, the experimental observation of $F_k$ happens much faster than the calculation of $P_k$ on a classical supercomputer, which constitutes the principle of quantum advantage. However, for a finite sample size, $F_k$ never coincides exactly with $P_k$, and the experimental result is accepted as a solution if it lies in the $\epsilon$-vicinity of the exact solution, where the distance
can be measured, e.g., by the Euclidean norm $\epsilon= \sqrt{\sum_k(F_k-P_k)^2}$ or some other measure. In addition, boson samplers are subject to losses and indistinguishability degradation, e.g., due to the multi-mode structure of squeezed light pulses \cite{Qi20,Shchesnovich22,Shi22}. Specifically, when the overall transmission $\eta$ of the boson sampler drops below a threshold value $\eta_\infty(r)$, the outcome probabilities can be efficiently calculated using a classical algorithm, eliminating the quantum advantage \cite{Qi20}.
The threshold transmission $\eta_\infty(r)$ can be reduced below the actual transmission $\eta$ by increasing the squeezing parameter $r$, making the exact classical simulation infeasible. However, an approximate simulation would still be possible: There would exist a fast classical algorithm to calculate a probability distribution $\tilde{P}_k$ that lies within the $\epsilon'$-vicinity of the exact solution $P_k$. For the quantum advantage to hold, it is necessary that $\epsilon \ll \epsilon'$, meaning the boson sampler produces results significantly closer to the exact solution than any classical simulation.
Increasing the squeezing parameter would eventually approach the limits of applicability of the parametric approximation, which would result in the growth of $\epsilon$, because the quantum model (in the parametric approximation) would become less precise. Consequently, approximation limits must be formulated in an error-specific manner as $V(r,\alpha,\epsilon)<1$, where $V(r,\alpha,\epsilon)$ is a real indicator function defining the validity region in the $(r,\alpha)$ space for a given acceptable error $\epsilon$. The Hillery-Zubairy limit can be rewritten in this form using the indicator function $V_{HZ}(r,\alpha,\epsilon)=\max(2,r)e^{2r}/2\alpha\epsilon$. However, as we demonstrate, this result holds only for sufficiently large values of $\epsilon$ (above $10^{-2}$), whereas modern boson samplers require much smaller acceptable error thresholds. 

In this paper, we introduce a new approach to the analytical calculation of the joint quantum state of pump and signal for the Hamiltonian given by Eq.~(\ref{Ham}). Unlike the parametric approximation, this solution is consistent with the field-energy conservation law, which is why we refer to it as ``isoenergetic.'' Moreover, it enables the calculation of various functions of interest for the signal and pump modes beyond the validity region of the parametric approximation and, more importantly, allows for the prediction of error-specific approximation limits.

The paper is structured as follows. In Sec.~\ref{sec:Model}, we establish the conditions under which the Hamiltonian (\ref{Ham}) applies to the single-pass generation of squeezed light, examine the general structure of the joint pump-signal state, and demonstrate how this structure is violated by the parametric approximation. In Sec.~\ref{sec:Pert}, we develop a perturbative approach to derive corrections to the mean number of signal photons and the variance of the squeezed quadrature. These corrections match those obtained via perturbative path integration \cite{Hillery84}. We also formulate the fundamental problem of the perturbative approach for unbounded operators.
A new approach is presented in Sec.~\ref{sec:Iso}, where we determine its domain of validity along with the revised validity domain of the parametric approximation. The analytical formulas are verified by numerical simulations with a moderate number of pump photons. The results are summarized in the Conclusion.

\section{Quantum model for degenerate single-mode downconversion \label{sec:Model}} 

\subsection{Spatial field evolution in single-pass downconversion \label{sec:Spatial}}

We consider a crystal of length $L$ illuminated by a pulsed pump beam at wavelength $\lambda_p$ polarized along one of the principal axes of the crystal. In the process of type-I (type-0) frequency-degenerate downconversion, a signal wave appears at wavelength $2\lambda_p$, it is polarized along another (the same) principal axis and propagates collinearly with the pump. The central frequencies of the waves are $\omega_p=2\pi c/\lambda_p$ with $c$ the speed of light in vacuum and $\omega_s=\omega_p/2$. The two waves are quasi-phase-matched by periodical poling with the poling period $\Lambda$. The positive-frequency part of the field of each wave is \cite{LoudonBook}
\begin{equation}\label{FourierSignal}
E_\mu^{(+)}(z,t) = i\mathcal{E}_\mu\int  \epsilon_\mu(z,\Omega)e^{ik_\mu(\Omega)z-i(\omega_\mu+\Omega)t} \frac{d\Omega}{2\pi},		
\end{equation}
where $\mu$ takes the values $\{p,s\}$ for the pump and signal waves, respectively, $t$ is time, $\Omega$ denotes the frequency detuning from the carrier frequency,  $k_\mu(\Omega)$ is the wave vector of the corresponding wave at frequency $\omega_\mu+\Omega$, and 
\begin{equation}
\mathcal{E}_\mu = \left(\frac{\hbar\omega_\mu}{2\varepsilon_0c\mathcal{A}n_\mu}\right)^{\frac12}
\end{equation}
with $\varepsilon_0$ the vacuum permittivity, $\mathcal{A}$ the cross-sectional area of the light beam, and $n_\mu$ the refractive index of the corresponding wave. The spectral amplitude of the pump or signal wave, $\epsilon_\mu(z,\Omega)$, is the annihilation operator of a photon at position $z$ with the frequency $\omega_\mu+\Omega$, satisfying the canonical equal-space commutation relations \cite{Huttner90,Kolobov99,Horoshko22} $\left[\epsilon_\mu(z,\Omega),\epsilon_\nu^\dagger(z,\Omega')\right]
= 2\pi\delta_{\mu\nu}\delta(\Omega-\Omega')$. The evolution of this operator along the crystal is described by the spatial Heisenberg equation \cite{Shen67,Caves87,Horoshko22,Kopylov24}
\begin{equation}\label{evolution}
    \frac{\partial}{\partial z}\epsilon_\mu(z,\Omega) = \frac{i}\hbar\left[\epsilon_\mu(z,\Omega),G(z)\right],
\end{equation}
where the spatial Hamiltonian $G(z)$ is given by the momentum transferred through the plane $z$ \cite{Horoshko22} and equals
\begin{equation}\label{G}
    G(z) = \chi(z)\int\limits_{-\infty}^{+\infty} E^{(+)}_p(z,t) E^{(-)}_s(z,t) E^{(-)}_s(z,t)dt + \text{H.c.},
\end{equation}
where $E^{(-)}_\mu(z,t)=E^{(+)\dagger}_\mu(z,t)$ is the negative-frequency part of the field and $\chi(z)=2\varepsilon_0\mathcal{A}d(z)$ is the coupling coefficient with $d(z)$ the second-order nonlinear susceptibility of the crystal. In a bulk crystal, $d(z)=d_\text{eff}$ is a constant. In a periodically poled crystal, $d(z)$ changes sign every distance of $\Lambda/2$, where $\Lambda$ is the poling period, i.e., represents a meander function. This function can be decomposed into Fourier series, where only one term, let it be one of the order $-1$, affects the phase matching \cite{BoydBook}. Thus, we write $d(z)\approx(2/\pi)d_\text{eff}\exp(-2\pi iz/\Lambda)$. Substituting Eqs.~(\ref{FourierSignal}) into Eq. (\ref{G}) and performing the integration, we obtain the spatial Hamiltonian in the form
\begin{eqnarray}\nonumber
G(z) &=& -i\hbar\gamma
\int\limits_{-\infty}^{+\infty} \frac{d\Omega}{2\pi}\frac{d\Omega'}{2\pi} \epsilon_p(z,\Omega+\Omega')\epsilon_s^\dagger(z,\Omega)\epsilon_s^\dagger(z,\Omega') \\\label{Gbis}
&\times& e^{i\Delta(\Omega,\Omega')z} + \text{H.c.},
\end{eqnarray}
where $\gamma=4\varepsilon_0\mathcal{A}d_\text{eff}\mathcal{E}_p\mathcal{E}_s^2/(\pi\hbar)$ is the coupling constant and $\Delta(\Omega,\Omega')=k_p(\Omega+\Omega')-k_s(\Omega)-k_s(\Omega')-2\pi/\Lambda$ is the phase mismatch for the two interacting waves.

The solution of Eq. (\ref{evolution}) has the form $\epsilon_s(z,\Omega)= U^\dagger\epsilon_s(0,\Omega)U$, where the evolution operator is $U=\mathcal{T}\exp\left[\frac{i}{\hbar}\int_0^L G(z)dz\right]$. Here, the symbol $\mathcal{T}$ denotes a space-ordering operator, putting the field operators with higher $z$-values to the left in the expansion of the exponential. It has been shown analytically for continuous-wave \cite{Lipfert18} and numerically for pulsed downconversion \cite{Christ13} that space ordering can be omitted when the degree of squeezing does not exceed 12 dB ($r\approx1.4$). For higher values of $r$, we need to calculate the space-ordering effects by evaluating the higher terms of the Magnus expansion \cite{Quesada14}, which is mathematically challenging.

The situation is much simpler in the quasi-monochromatic limit, where the pump and signal fields have narrow bandwidths $\Delta\omega_p=2\pi\Delta f_p$ and $\Delta\omega_s=2\pi\Delta f_s$, respectively. In this case, we write $E_p^{(+)}(z,t)=i\mathcal{E}_p\sqrt{\Delta f_p}a(z) e^{ik_p(0) z-i\omega_p t}$, $E_s^{(+)}(z,t)=i\mathcal{E}_s\sqrt{\Delta f_s}b(z) e^{ik_s(0) z-i\omega_s t}$, where $a$ and $b$ are photon annihilation operators of the corresponding waves defined as 
\begin{equation}
a(z) =  \frac1{\sqrt{\Delta f_p}}\int_{\Delta\omega_p}\epsilon_p(z,\Omega) \frac{d\Omega}{2\pi} 
\end{equation}
and similarly for $b(z)$, which results in commutators $\left[a(z),a^\dagger(z)\right]=\left[b(z),b^\dagger(z)\right]=1$. We imply that both quasi-monochromatic waves have a duration $t$ and are perfectly phase matched, i.e., $k_p(0)-2k_s(0)=2\pi/\Lambda$. In this case, Eq. (\ref{G}) transforms into 
\begin{equation}\label{Gter}
G = -i\hbar\gamma\sqrt{\Delta f_p}\Delta f_s t\left(ab^{\dagger2}-a^\dagger b^2\right) = -Ht/L,
\end{equation}
where $H$ is given by Eq. (\ref{Ham}) with $L$ the crystal length and $\kappa=\gamma\sqrt{\Delta f_p}\Delta f_s L$. The evolution operator in the quasi-monochromatic limit reads 
\begin{equation}
U = e^{iGL/\hbar} = e^{-iHt/\hbar},
\end{equation}
i.e., it coincides with that of two cavity modes interacting with the Hamiltonian $H$ during time $t$.

\subsection{Invariant subspaces of the Hamiltonian}

We are interested in the evolution of the initial state $|\Phi\rangle|0\rangle$ with an arbitrary state of the pump $|\Phi\rangle$ in the strong pump limit (i.e., for a large average number of photons and relatively small dispersion)  and vacuum in the signal mode.  Our approach is based on the fact that the Hamiltonian $H$ in Eq. (\ref{Ham}) conserves the total optical energy $\hbar\omega_p\hat N$, where
\be
\hat{N} \equiv  {a}^\dag {a}+\frac12  {b}^\dag {b}.
\en{energy}
Since $[ {H}, \hat{N}]=0$, the Hilbert space $\mathcal{H}$ decomposes into a direct sum of orthogonal invariant subspaces
\be
\mathcal{H} =  \mathcal{H}_0 \oplus \mathcal{H}_1 \oplus \mathcal{H}_2 \oplus \ldots  \oplus \mathcal{H}_N\oplus \ldots, \en{decomp}
with the subspace $\mathcal{H}_N$  corresponding to the eigenstates of $ \hat{N}$, Eq.~(\ref{energy}), with the (integer) eigenvalue $N\ge 0$.  
Therefore, we need to study the evolution of the initial state $|N\rangle|0\rangle$ in the orthogonal subspace $\mathcal{H}_N$, where $|N\rangle  \equiv \frac{\left( {a}^\dag\right)^{N} }{\sqrt{N! }}|0\rangle$ is the Fock state of the pump. Our main goal is to find the state  
\be
|\Psi^{(N)}\rangle\equiv e^{-\frac{it}{\hbar} {H}}|N\rangle|0\rangle = \sum_{n=0}^N \Psi^{(N)}_{n}|N-n\rangle|2n\rangle,
\en{Amp}
where $|2n\rangle = \frac{\left( {b}^\dag\right)^{2n}}{\sqrt{ (2n)!}}|0\rangle$ is the Fock state of the signal. In the following, we will use the dimensionless time $\tau = \kappa t$.

\subsection{Strong coherent pump}  
In the strong pump regime only the subspaces $\mathcal{H}_N$ with large $N\gg1$ contribute significantly to the input state. The main application is the case of a coherent state of the pump    
\be
 |\alpha\rangle = e^{-\frac{\alpha^2}{2}}\sum_{N=0}^\infty \frac{\alpha^N}{\sqrt{N!}}|N\rangle
\en{CohSt}
(we can always set $\alpha>0$ by the $SU(1)$-invariance of the quantum states) with a large average number of photons $\langle \hat N\rangle =\alpha^2\gg 1$, as is usually assumed in the standard parametric approximation \cite{Loudon87,Lvovsky15}. In this case, the probability concentration inequality \cite{Hoeffding94} applied to the Poisson distribution of photon counts from the coherent state $|\alpha\rangle$ gives 
\be
\mathrm{Prob}\left(\left|\frac{N}{\alpha^2} - 1\right|\ge  \frac{c}{\alpha} \right)\le 2\exp\left(-\frac{c^2}{2}\right),
\en{bound1}
for arbitrary $0\le c< \alpha$. Therefore, to obtain an error $\epsilon$ in probability, the index of a contributing subspace $\mathcal{H}_N$ in Eq. (\ref{decomp}) must belong to the $\epsilon$-confidence interval
\be
\Omega_\epsilon(\alpha): \left\{N, \left|\frac{N}{\alpha^2} - 1\right| \le   \frac{\sqrt{2\ln(2/\epsilon)}}{\alpha}\right\}.
\en{bound}

\subsection{Energy conservation and  the parametric approach}

Before proceeding with our solution of the Schrödinger equation, let us recall the standard parametric approach, which assumes a strong coherent pump with amplitude $\alpha \gg 1$, replacing the pump boson operator (in the interaction picture) with a scalar parameter, ${a} \to \alpha$. This procedure results in a Gaussian squeezed state within the parametric approximation \cite{Loudon87,Lvovsky15}:
\begin{eqnarray}\label{sq}
|r\rangle = \sqrt{\sech  r}\sum_{n=0}^\infty \binom{2n}{n}^\frac12\frac{\tanh^n r}{2^n}|2n\rangle, \,\,   r\equiv 2\alpha \tau.
\end{eqnarray}
Since the coherent pump state remains unaffected in the parametric approximation, the correct normalization of the Gaussian state in Eq.~(\ref{sq}) must arise from compensating positive and negative (unphysical) corrections to the norm of the combined pump-signal state in invariant subspaces $\mathcal{H}_N$.
To see this, we project the joint state $|\alpha\rangle |r\rangle$ onto the invariant subspaces $\mathcal{H}_N$. Introducing the probability of detecting $M$ photons from the coherent state $\mathcal{P}_M   = e^{-\alpha^2} \alpha^{2M}/M!$, and $2n$ from the Gaussian squeezed state, $p_{2n} = \sech r\binom{2n}{n}(\frac12\tanh r)^{2n}$,  collecting the factors according to $N = M+n$ (here  the number of photons in the pump mode is $M$, whereas $N$ is the total number of photons), we have 
\begin{eqnarray}\nonumber
|\alpha\rangle|r\rangle &=& \sum_{M=0}^\infty \sqrt{\mathcal{P}_M}|M\rangle \sum_{n=0}^\infty \sqrt{p_{2n}}|2n\rangle\\\label{EqA1}
&=&  \sum_{N=0}^\infty \sqrt{\mathcal{P}_N}  \sum_{n=0}^N \frac{\sqrt{(N)_n}}{\alpha^n}\sqrt{p_{2n}} |N-n,2n\rangle\\\nonumber
&=& \sum_{N=0}^\infty \sqrt{\mathcal{P}_N}|\widetilde{\Psi}^{(N)}\rangle,
\end{eqnarray}
where $(N)_n = N!/(N-n)!$ and $|\widetilde{\Psi}^{(N)}\rangle$ is the state in the orthogonal subspace $\mathcal{H}_N$, Eq.~(\ref{Amp}), with $ \Psi^{(N)}_n$ (observe that by definition  $\widetilde{\Psi}^{(N)}_n=0$ for $n> N$) substituted for 
\be
\widetilde{\Psi}^{(N)}_n  = \sqrt{\sech r} \binom{2n}{n}^\frac12  \frac{ \sqrt{(N)_n}}{\alpha^n} \left(\frac{\tanh r}{2 }\right)^n,
 \en{PsiG}
which can be understood as the state resulting from the $N$-photon component of the pump. The norm of the state $|\widetilde{\Psi}^{(N)}\rangle$ is found in Appendix \ref{sec:AppendixPA}.  Up to the first order in $1/\alpha^2$, it  reads
\be
 \langle\widetilde{\Psi}^{(N)}|\widetilde{\Psi}^{(N)}\rangle \approx 1 + \frac{\sinh^2 r}{2} \left(\frac{N}{\alpha^2}-1\right).  
\en{EqA5}
Whereas, the identity $\sum_N\mathcal{P}_N\langle\widetilde{\Psi}^{(N)}|\widetilde{\Psi}^{(N)}\rangle=1$, which follows from Eq. (\ref{EqA1}), guarantees that the parametric approach maintains the correctly normalized joint system state at all times, Eq.~(\ref{EqA5}) shows that there are mutually compensating non-physical corrections to the norm of the quantum state projection onto the invariant subspaces $\mathcal{H}_N$: negative for $N < \alpha^2$ and positive for $N > \alpha^2$. We will use this and similar facts to derive the error-specific bound on the parametric approximation and on a more precise approximation derived below.

\section{Perturbative  solution and the parametric approximation \label{sec:Pert}}

\subsection{Decomposition in the inverse pump amplitude \label{sec:Decomposition}}
We consider a typical regime in which the pump contains much more photons at the input than the signal at the output, e.g. $\alpha\approx2\cdot10^6$ and $\langle b^\dagger b\rangle\approx 10$, as in the boson sampling experiment \cite{Zhong20}. We write the state of the pump and signal as
\begin{equation}\label{Psiab}
\left|\Psi\right\rangle_{ab} = e^{-\frac{i}\hbar  {H}t}|\alpha\rangle_a|0\rangle_b,
\end{equation}
where $H$ is given by Eq. (\ref{Ham}). The initial coherent state of the pump can be represented as $|\alpha\rangle=D(\alpha)|0\rangle$, where $D(\alpha)=\exp\left(\alpha a^\dagger-\alpha a\right)$ is the shift operator for $\alpha\in\mathbb{R}$. Then, the state of the pump and signal, Eq. (\ref{Psiab}), can be rewritten as $\left|\Psi\right\rangle_{ab} = D(\alpha)U(r)|0\rangle_a|0\rangle_b$, where 
\begin{equation}\label{Ur}
U(r) = e^{\frac{r}2\left[(1+\nu a)b^{\dagger 2}-(1+\nu a^\dagger) b^2\right]}
\end{equation}
is a unitary operator with $r=2\alpha\kappa t$ and $\nu=1/\alpha$. Since in a typical experiment, $\nu$ is of the order of $10^{-6}$, we can look for a perturbative solution for the evolution operator in orders of $\nu$ at a fixed $r$. Differentiating Eq. (\ref{Ur}) over $r$, we obtain
\begin{equation}\label{Ureq}
\frac{dU(r)}{dr} =\frac12\left[(1+\nu a)b^{\dagger 2}-(1+\nu a^\dagger) b^2\right]U(r).
\end{equation}
Now, as usual in perturbation theory, we consider the part linear in $\nu$ as a perturbation and write $U(r)=S(-r)U_I(r)$, where $S(r)=\exp\left[\frac{r}2(b^2-b^{\dagger 2})\right]$ is the squeezing operator and $U_I(r)$ is the evolution operator in the interaction picture, satisfying the equation
\begin{eqnarray}\label{UreqI}
\frac{dU_I(r)}{dr} &=&-iH_I(r)U_I(r),\\
H_I(r )&=& \frac{i\nu}2\left[a\left(b^{\dagger}\cosh{r}+b\sinh{r}\right)^2\right.\\\nonumber
&-&\left.a^\dagger \left(b\cosh{r}+b^{\dagger}\sinh{r}\right)^2\right],
\end{eqnarray}
where we have taken into account that the squeezing operator realizes a Bogoliubov transformation of the signal mode operators: $S(r)bS(-r)=b\cosh{r}+b^{\dagger}\sinh{r}$. A similar interaction Hamiltonian was recently obtained in an approach aiming to separate Gaussian and non-Gaussian evolution in squeezed light generation \cite{Yanagimoto22}. In contrast, we separate only the initial coherent state of the pump, and, in our formalism, the argument of the shift operator $D(\alpha)$ is a constant. The solution of Eq. (\ref{UreqI}) has the form
\begin{equation}\label{UIr}
U_I(r) = \mathcal{T} \exp\left(-i\int\limits_0^r H_I(r)dr\right),
\end{equation}
where $\mathcal{T}$ is the time-ordering operator placing the operators $H_I(r)$ with higher $r$ to the left in the Taylor decomposition of the exponential. Thus, the state of the pump and signal, Eq. (\ref{Psiab}), can be rewritten as
\begin{equation}\label{Psiab2}
\left|\Psi\right\rangle_{ab} = D(\alpha)S(-r)U_I(r)|0\rangle_a|0\rangle_b.
\end{equation}
Note that the entanglement between the pump and the signal is created by the operator $U_I(r)$, since $D(\alpha)$ only acts on the pump mode and $S(-r)$ only acts on the signal mode. 

In zeroth order in $\nu$, $U_I(r)=1$ and the state of the two modes is $\left|\Psi\right\rangle_{ab} = |\alpha\rangle_a|r\rangle_b$, where $|r\rangle=S(-r)|0\rangle$ is the squeezed state, defined in Eq.~(\ref{sq}). Thus, the zeroth order corresponds to the parametric approximation.

For calculating averages of signal mode operators up to the first nonvanishing correction, we need to keep the first three terms in the Taylor decomposition of the exponential in Eq. (\ref{UIr}). The second-order (in $\nu$) term of $U_I(r)|0\rangle_a|0\rangle_b$ contains components $|0\rangle_a$ and $|2\rangle_a$. The latter is orthogonal to the lower-order components and can be discarded, but the former interferes with the zero-order contribution and should be kept. In this way, we obtain
\begin{equation}\label{UIrapprox}
U_I(r)|0\rangle_a|0\rangle_b \approx 
\left(1- \frac{\nu^2}{32} B_0^\dagger - \frac{\nu}4 a^\dagger B_1^\dagger\right)|0\rangle_a|0\rangle_b,
\end{equation}
where $B_1^\dagger=s^2+b^{\dagger 2}(cs-r)$, $B_0^\dagger=g_0+g_2b^{\dagger 2}+g_4b^{\dagger 4}$, and we introduce the following shortcuts: $c=\cosh{r}$, $s=\sinh{r}$, $g_0=3s^4+2s^2-4rcs+2r^2$, $g_2=6cs^3+8cs-10rs^2-8r$, and $g_4=s^4+3s^2-2rcs-r^2$. Note that Eq.~(\ref{UIrapprox}) preserves the norm in the second order of $\nu$, that is, ${}_{ab}\left\langle\Psi\right| \left.\Psi\right\rangle_{ab} =1+\mathcal{O}(\nu^4)$.

Equations~(\ref{Psiab2}) and (\ref{UIrapprox}) allow us to calculate the corrections to the mean photon number in the signal mode $\hat n=b^\dagger b$ and the variance of the squeezed signal quadrature $X_-=-i(b-b^\dagger)/2$ in the first nonvanishing (second) order of $\nu$:
\begin{eqnarray}\label{corrMean}
\langle\hat n\rangle &=& {}_{ab}\left\langle\Psi\right| b^\dagger b\left|\Psi\right\rangle_{ab} = s^2\left[1+\nu^2f_\text{mean}(r)\right],\\\label{corrSq}
\langle\Delta X_-^2\rangle &=& {}_{ab}\left\langle\Psi\right| X_-^2\left|\Psi\right\rangle_{ab} = \frac14 e^{-2r}\left[1+\nu^2f_\text{sq}(r)\right],
\end{eqnarray}
where $\Delta X_-=X_- -\langle X_-\rangle$ and
\begin{eqnarray}
f_\text{mean}(r) &=& \frac14\left[r^2(2+s^{-2})+2rcs^{-1}-3s^2-3\right],\\
f_\text{sq}(r) &=& \frac12\left[r^2-2r(s^2+cs+1)+s^4+cs^3 \right.\\\nonumber
&&\left.+s^2+2cs\right].
\end{eqnarray}
Equations~(\ref{corrMean}) and (\ref{corrSq}) coincide with those obtained by Hillery and Zubairy by a perturbative path integration technique \cite{Hillery84}, allowing us to conclude that the two approaches are equivalent. 

Two features of the perturbative solution can be distinguished in the very high gain regime, where $s\gg r$ (which occurs at $r>4.5$). First, the mean number of photons grows as $\langle\hat n\rangle_0=s^2$ in the parametric approximation, while the perturbative correction results in a lower number of photons, since $f_\text{mean}(r)\approx-0.75s^2$. Second, the variance of the squeezed quadrature decreases as $e^{-2r}/4$ in the parametric approximation, while the perturbative correction results in its growth and eventual saturation \cite{Hillery84,Crouch88} at $r=0.5\ln(4\alpha)$, since $f_\text{sq}(r)\approx e^{4r}/16$. Similar features were obtained for a perturbative solution to the closely related problem of field evolution in a nondegenerate parametric amplifier with a \emph{trilinear} Hamiltonian \cite{Xing23}. These features will be compared with those of the non-perturbative solution found in Sec.~\ref{sec:Iso}.

\subsection{Domain of validity of the parametric approximation via the perturbative solution}

\begin{figure}[!ht]
\centering
\includegraphics[width=\linewidth]{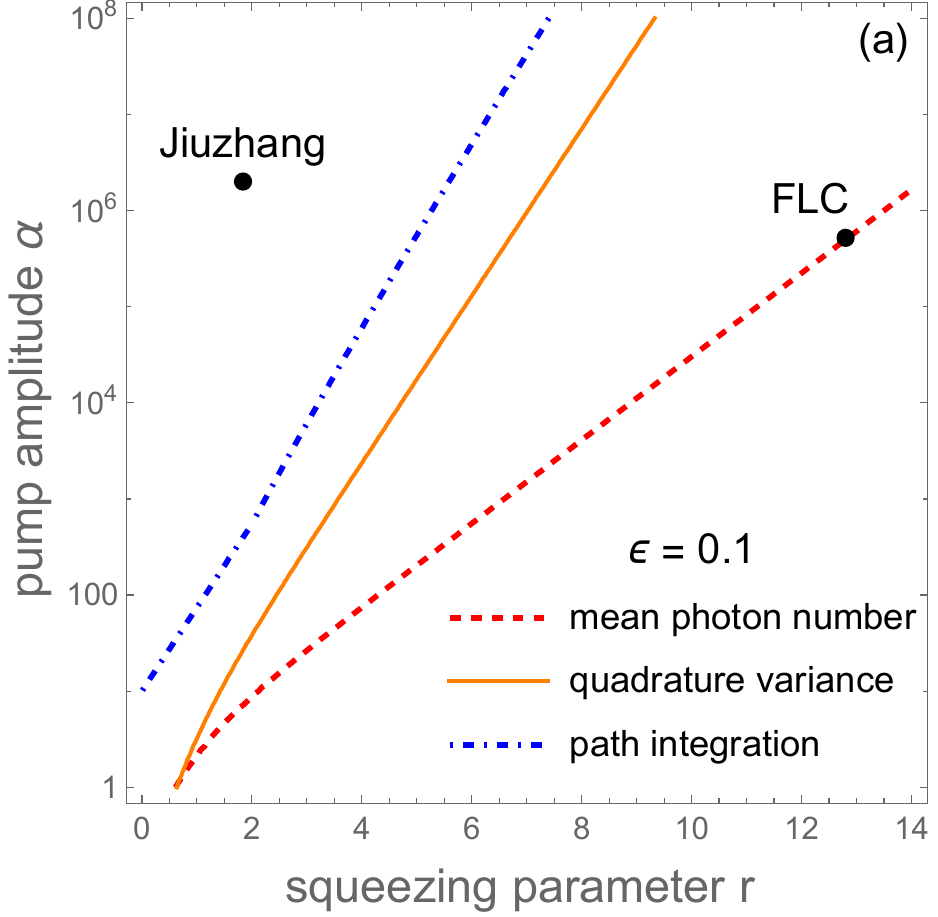}
\includegraphics[width=\linewidth]{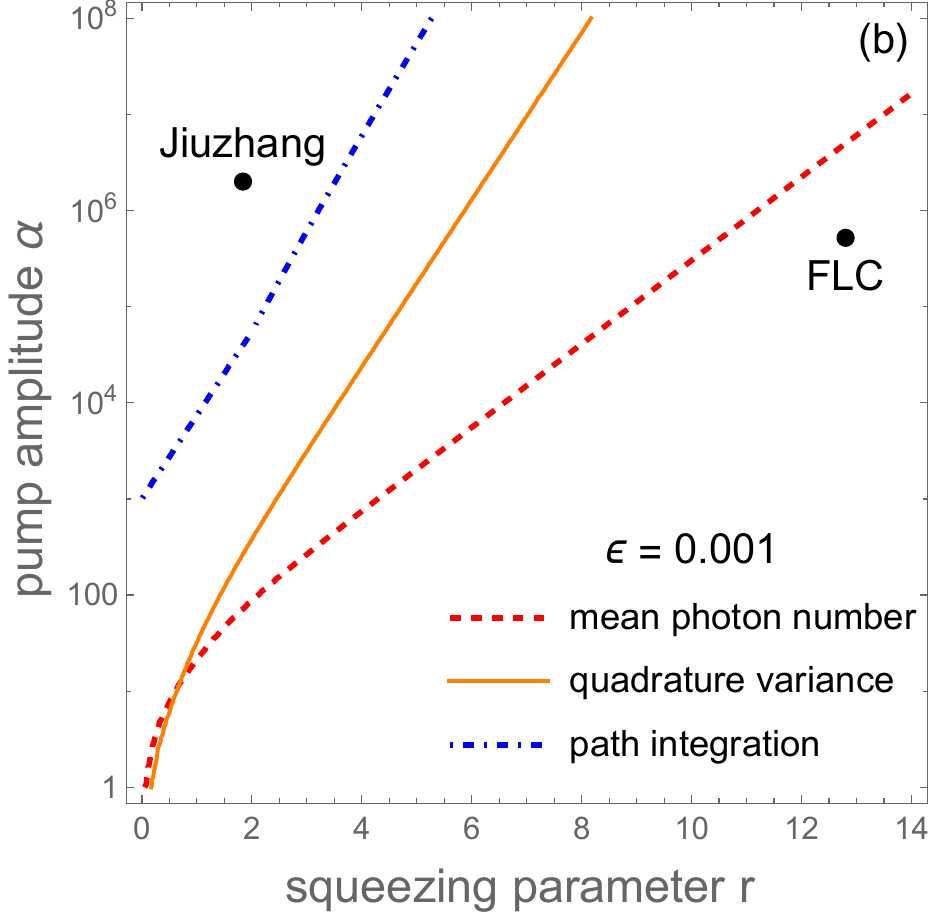}
\caption{Domains of validity of the parametric approximation. The area above the red dashed line corresponds to  $V_\text{mean}(r,\alpha,\epsilon)<1$; in this area, the mean number of signal photons given by the parametric approximation has an error below $\epsilon$. The area above the orange solid line corresponds to  $V_\text{sq}(r,\alpha,\epsilon)<1$; in this area, the squeezed quadrature variance given by the parametric approximation has an error below $\epsilon$. The area above the blue dash-dotted line corresponds to  $V_{HZ}(r,\alpha,\epsilon)<1$; in this area, any average of the field operators given by the parametric approximation has an error below $\epsilon$. The two points correspond to the experimental conditions of the Jiuzhang boson sampler \cite{Zhong20} and to the FLC experiment \cite{Florez20}. \label{fig:ValidityPert}}
\end{figure}

By requiring the corrections provided by Eqs.~(\ref{corrMean}) and (\ref{corrSq}) be less than $\epsilon$, we obtain the limitations of the parametric values of the mean photon number and the squeezed quadrature variance in error-specific form $V(r,\alpha,\epsilon)<1$, where $V_\text{mean}(r,\alpha,\epsilon)=\nu^2 |f_\text{mean}(r)|/\epsilon$ and, similarly, $V_\text{sq}(r,\alpha,\epsilon)=\nu^2 |f_\text{sq}(r)|/\epsilon$. The validity domains of the two approximations are compared in Fig.~\ref{fig:ValidityPert} for various values of acceptable error $\epsilon$. We also show the validity domain $V_{HZ}(r,\alpha,\epsilon)<1$, obtained in Ref.~\cite{Hillery84} and discussed in the Introduction.

In addition, two points show the parameters of two experiments with high degrees of squeezing. The first of them corresponds to the Jiuzhang boson sampler \cite{Zhong20}, where the maximum squeezing parameter was $r=1.84$ while the pump pulse energy was $U\approx1$~$\mu J$, which corresponds to the mean number of $4\cdot10^{12}$ photons at a wavelength of 776 nm or to $\alpha=2\cdot10^{6}$. The second corresponds to the experiment of Fl\'orez, Lundeen and Chekhova (FLC) \cite{Florez20}, where a squeezing parameter $r=12.8$ was reached at the pump pulse energy $U\approx$~0.1 $\mu J$, which corresponds to the mean number of $2.7\cdot10^{11}$ photons at the wavelength of 532 nm or to $\alpha=5.2\cdot10^{5}$. Up to this value of the pump amplitude, the authors of Ref.~\cite{Florez20} observed a growth of the mean number of signal photons according to the parametric law $\langle\hat n\rangle_0=\sinh^2(r)$, which broke at higher values of the pump amplitude. As we see in Fig. \ref{fig:ValidityPert}(a), this breaking point lies exactly on the border of the validity region of the parametric value for $\langle n\rangle$ at $\epsilon=0.1$, which is a typical value of the ``negligible error'' in a physical experiment.

We also see in Fig. \ref{fig:ValidityPert} that the region of validity of the parametric formula for $\langle \Delta X_-^2\rangle$ is very different from that for $\langle \hat n\rangle$. Indeed, we obtain from Eqs.~(\ref{corrMean}) and (\ref{corrSq}) in the limit of high $r$, that the corrections to $\langle\hat n\rangle$ and $\langle \Delta X_-^2\rangle$ have different scales: $\nu^2 f_\text{mean}(r)\to-0.75\langle\hat n\rangle_0/\alpha^2$ and $\nu^2 f_\text{sq}(r)\to\langle\hat n\rangle_0^2/\alpha^2$, that is, the latter grows faster with $\langle\hat n\rangle_0$. For other operator averages, the growth of the error with $\langle\hat n\rangle_0$ may be even faster. The ultimate limit of validity of the parametric approximation for all possible operator averages is set by the inequality $V_{HZ}(r,\alpha,\epsilon)<1$ following from the analysis of perturbative path integration in Ref.~\cite{Hillery84}.

\subsection{Problem with the perturbative solution for unbounded operators}
The perturbative approximation made in the previous sections relies on a key assumption: If the lowest-order corrections are small, then the higher-order corrections must be even smaller. In other words, we assume that the perturbative series in the parameter $\nu$ is convergent.
However, when unbounded-norm operators are present, they can lead to an unbounded growth of successive terms in the perturbative expansion. In such cases, the series is known as ``asymptotic.'' A fundamental property of asymptotic series (see, e.g., Ref.~\cite{Olver}) is that they can be approximated up to an optimal error, achieved by truncating the series at an optimal number of terms, often high above the lowest-order correction.

In our case, the perturbation $H_I$ in Eq.~(\ref{UreqI}) contains the unbounded-norm operator $a^{\dag} b^{\dag2}$, which acts on the vacuum state of the system and, as a result, the higher orders of the perturbative expansion can grow uncontrollably. Consequently, the error (and therefore the validity domain) of the perturbative approach cannot be reliably estimated using the perturbation theory alone. To determine when the perturbative approach remains valid, we need an alternative method that avoids unbounded-norm operators. This can be achieved by solving the Schrödinger equation within the invariant \textit{finite-dimensional} subspaces $\mathcal{H}_N$ of the Hilbert space. This approach respects the conservation of optical energy, as these invariant subspaces correspond to well-defined total energy values of the optical modes. For this reason, we refer to it as the ``isoenergetic approach" and call the resultant model also ``isoenergetic.''

\section{Isoenergetic approach \label{sec:Iso}}

In the strong pump regime, $N \gg 1$, we propose an analytical solution to the Schrödinger equation in the Fock space within all invariant subspaces $\mathcal{H}_N$ that satisfy the bound in Eq.~(\ref{bound}) for a given relative error $\epsilon$. This solution accurately describes the quantum amplitudes $\Psi^{(N)}_n$ in Eq.~(\ref{Amp}) for $n \leq n(N, \epsilon)$, where $n(N, \epsilon)$ represents a very large number of signal photon pairs when $N \gg 1$.

Our approach is fully consistent with optical energy conservation as described by Eq.~(\ref{energy}), ensuring that the norm of each projected state within its respective invariant subspace is preserved. This prohibits any unphysical transfer of optical energy between different invariant subspaces of the Hilbert space. In contrast, as Eq.~(\ref{EqA5}) demonstrates, the parametric approximation \textit{relies} on such unphysical energy transfer, introducing positive norm corrections in some invariant subspaces and negative ones in others.

Moreover, the isoenergetic approach allows us to precisely analyze the error behavior of the parametric approximation and determine its validity domain in the parameter space $(r, \alpha)$. Specifically, we identify the regime where perturbative corrections in the strong pump limit ($\alpha \gg 1$) can be neglected. We will show that, for sufficiently strong pumps, the error of the parametric approximation grows rapidly upon crossing the validity boundary in the $(r, \alpha)$ plane. Furthermore, we derive a more refined approximation that, at the cost of a higher relative error, reproduces the parametric one while possessing a broader domain of validity in $(r, \alpha)$.

\subsection{Approximate solution to the Schr\"odinger equation}

The amplitudes $\Psi^{(N)}_n(\tau)$ in Eq. (\ref{Amp}) satisfy the following Schr\"odinger equation in the Fock space   
 \be
 \frac{\rd  \Psi^{(N)}_n}{\rd \tau} = \sqrt{\beta_{n-1}} \Psi^{(N)}_{n-1} - \sqrt{\beta_{n}} \Psi^{(N)}_{n+1},  \,\,  \Psi^{(N)}_n (0)=\delta_{n,0},
 \en{Schr} 
where $\beta_n = (N-n)(2n+1)(2n+2)$, with the boundary condition: $\Psi^{(N)}_{-1}(\tau) =\Psi^{(N)}_{N+1}(\tau)=0$.  Equation~(\ref{Schr}) follows from the application of the Hamiltonian in Eq. (\ref{Ham}), using the commutation relations for the boson operators and observing that $(N)_n(2n)!=\prod_{k=0}^{n-1}\beta_k $. An equivalent form of Eq. (\ref{Schr}), useful for analytical analysis, is given for a rescaled amplitude. Setting 
 \be 
 \Psi^{(N)}_n = \left(\prod_{k=0}^{n-1}{\beta_k}\right)^\frac12 \gamma^{(N)}_n,
 \en{Psigam} 
we obtain
   \be
 \frac{\rd  \gamma^{(N)}_n}{\rd \tau} = \gamma^{(N)}_{n-1} -  \beta_{n} \gamma^{(N)}_{n+1},\quad \gamma^{(N)}_n(0)=\delta_{n,0}. 
 \en{Schr2} 
Although Eqs.~(\ref{Schr})-(\ref{Schr2}) with the initial condition $\gamma^{(N)}_n(0) = \delta_{n,0}$ admit an exact solution in the form of an infinite series \cite{Shchesnovich24}, we seek a simpler approximate solution valid for $N \gg 1$ and $n \ll \sqrt{N}$ (as discussed below). To achieve this, we leverage the fact that Eq.~(\ref{Schr2}) describes the propagation of an initial excitation at $n=0$. Thus, for sufficiently short evolution times (or equivalently, small squeezing parameter $r$), it is reasonable to assume that only the amplitudes $\gamma^{(N)}_n$ with $n \ll N$ contribute significantly to the solution.

An approximate model can be obtained by removing the small term on the order $\mathcal{O}(n/N)$ from $\beta_n$, that is, by considering Eqs. (\ref{Schr})-(\ref{Schr2}) with an approximate $\beta^{(a)}_n\equiv N(2n+1)(2n+2)$ (in this way we conserve the unitarity of evolution of the quantum amplitudes $\Psi^{(N)}_n$ in Eq. (\ref{Psigam})). In the following, we work mainly in the subspace $\mathcal{H}_N$, thus all parameters depend on the index of the subspace $N$, which we omit for simplicity of presentation. Introducing the parameter $r_N \equiv 2\sqrt{N}\tau$ and setting  
 \be
 \gamma^{(a)}_n (r)= \frac{f^{(a)}_n(r)}{n!(2\sqrt{N})^n},
\en{gamfApp} 
we obtain the following equation 
 \be
 \frac{\rd f^{(a)}_n}{\rd r} =n f^{(a)}_{n-1} -   \left(n+\frac12\right) f^{(a)}_{n+1},\,\, f^{(a)}_n(0)=\delta_{n,0}. 
 \en{Eqfn}
One can think of the approximate model in Eq. (\ref{Eqfn})  for the unbounded $n\ge0$ (for a finite $r$) as the mathematical limit as $N\to\infty$ of the original model in Eqs. (\ref{Schr})-(\ref{Schr2}). In this limit Eq. (\ref{Eqfn}) admits an exact solution. Setting   $f^{(a)}_n =  f^{(a)}_0T^n(r)$   and collecting  the terms with the factor ``$n$" in Eq. (\ref{Eqfn}), we obtain $\frac{\rd T}{\rd r} = 1-  T^2$ and $T(0)=0$, giving $T(r) = \tanh r$, whereas the remaining terms give $\frac{\rd f^{(a)}_0}{\rd r} =- \frac12 Tf^{(a)}_{0}$ with the solution $f^{(a)}_0 = \sqrt{\sech r}$.  

To see the order of the approximation error of the model in Eq. (\ref{Eqfn}) committed by adopting the approximate $\beta^{(a)}_n\equiv N(2n+1)(2n+2)$ and infinite number of amplitudes $n\ge 0$, we assume that the exact $\gamma_n$, which satisfies the exact Eq. (\ref{Schr2}), is related to the exact $f_n$ by the same Eq. (\ref{gamfApp}) and set $f_n =f^{(a)}_n+ \chi_n$. From  Eq.~(\ref{Schr2}) we get a linear equation with the source term for $\chi_n$
\begin{eqnarray}
\label{Eq4R}
 \frac{\rd \chi_n}{\rd r} &= & n\chi_{n-1}  -  \left(1-\frac{n}{N}\right) \left(n+\frac12\right) \chi_{n+1} \nonumber\\
&+&\frac{n}{N}\left(n+\frac12\right)  f^{(a)}_{n+1}(r),
\end{eqnarray}
with the  initial condition $\chi_n(0)=0$. By analyzing the integral version of Eq. (\ref{Eq4R}), we obtain 
the relative order of the correction term $\chi_n/f^{(a)}_n = \mathcal{O}[r(n^2+1)/N]$. An additional relative error on the same order appears when we compute the normalized Fock state amplitude by Eq. (\ref{Psigam}), using the approximate $\beta^{(a)}_k$-factors (the same factors as in Eq. (\ref{Eqfn})), for $0\le k\le n-1$, each having the relative error $\mathcal{O}(k/N)$, totaling to $\mathcal{O}[n(n+1)/N]$. Collecting the orders of the two errors we conclude that for $N\gg 1$  Eqs. (\ref{Schr}) admits the following approximate solution (in the subspace $\mathcal{H}_N$; we also show the dependence of $r$ on $N$ explicitly)
\begin{eqnarray}\label{Sol}
\Psi^{(N)}_n &=& \sqrt{\sech r_N}\binom{2n}{n}^\frac12\left(\frac{\tanh r_N}{2}\right)^n  \nonumber\\
& \times& \left\{1+ \mathcal{O}\left(\frac{(1+r_N)(n^2+1)}{N}\right)\right\}.
\end{eqnarray} 
Below we consider $r_N=\mathcal{O}(1)$, thus the order of the error is defined by $n(n+1)/N$. We see that only the amplitudes up to $n\le  \sqrt{\epsilon N}$ (from the set $0\le n\le N$) can be approximated to a given relative error of order $ \epsilon$. We call the solution defined by Eqs.~(\ref{Amp}) and (\ref{Sol}) isoenergetic approximation, because it slightly redistributes the relative weights of the components $|N-n\rangle|2n\rangle$ in the two-mode state, Eq.~(\ref{Amp}), but does not change the total energy, in contrast to the perturbative solution, considered in Sec.~\ref{sec:Pert}.   

\subsection{Comparison with the parametric approximation} 

Let us compare the leading order of the amplitudes in Eq.~(\ref{Sol}) with the corresponding amplitudes in the parametric approximation in Eq.~(\ref{PsiG}). There are two differences: (i) in the parametric approximation, the amplitudes $\widetilde{\Psi}_n^{(N)}$ of Eq.~(\ref{PsiG}) have a uniform parameter $r=2\alpha \tau$, while in the approximation of Eq. (\ref{Sol}), we have $r_N=2\sqrt{N}\tau$ in each subspace $\mathcal{H}_N$, (ii) there is an additional factor $\sqrt{(N)_n}/\alpha^n$ in the parametric approximation. Both differences are responsible for the higher-order error of the parametric approximation in comparison to the solution of Eq. (\ref{Sol}). Equation (\ref{Sol}) for $n\le \sqrt{\epsilon N}$ approximates the exact solution of Eq. (\ref{Schr2}) to the relative error on the order $\mathcal{O}( {\epsilon})$, while the state of Eq.~(\ref{PsiG}) can have a relative error of at least $\mathcal{O}(\sqrt{\epsilon})$. In order to show this, let us reduce to the uniform squeezing parameter $r$ in $\mathcal{H}_N$ by using the expansion $\tanh^n r_N = \tanh^n r \left[ 1 + \frac{2n}{\sinh(2r)}(r_N-r)+\ldots \right]$. For $N\in \Omega_\epsilon(\alpha)$ of Eq. (\ref{bound}) we have $|r_N-r| \le  r \sqrt{2\ln(2/\epsilon)}/\alpha$ and the relative error becomes 
\begin{eqnarray}
\label{exp_tanhr}
\tanh^n r_N &= &\tanh^n r\left[ 1 + \mathcal{O}\left(e^{-2r}n(r_N-r)\right) \right] \nonumber\\
& = &  \tanh^n r \left[ 1 + \mathcal{O}\left(e^{-2r}\frac{n}{\alpha} \right) \right] .
\end{eqnarray}
We get $\mathcal{O}(e^{-2r}n/\alpha) =\mathcal{O}(e^{-2r}\sqrt{\epsilon})$ in Eq. (\ref{Sol}) for $n\le n_N = \sqrt{\epsilon N}$. Hence, one of the sources of a higher error of the Gaussian squeezed state is the uniform squeezing parameter $r$.  Another source of a higher relative error is the extra factor in $ {\sqrt{(N)_n}}/{\alpha^n}$. For the latter, we can use the asymptotic estimate \cite{Shchesnovich21} 
 \be
 (N)_n =  N^n \exp\left( -\frac{n^2}{2N}\right)\left[1 + \mathcal{O}\left(\frac{n^3}{N^2}\right)\right].
 \en{factor}
In the confidence interval of Eq. (\ref{bound}) to the leading order in $\epsilon$ we get (for  $n\le n_N$, i.e., $n/\alpha = \sqrt{\epsilon}$)
\be
 {\sqrt{(N)_n}}/{\alpha^n}  = 1+ \mathcal{O}\left( \sqrt{\ln(1/\epsilon)}\frac{n}{\alpha}\right) = 1+\mathcal{O}(\sqrt{\epsilon}).         
\en{err2}

Therefore, the parametric approximation is less precise than the isoenergetic one because the relative error of the former is the square root of the relative error of the latter. We note that the same order of the relative error of the parametric approximation, i.e., $\mathcal{O}(\sqrt{\epsilon})=\mathcal{O}(n/\alpha)$, is found by comparison of the few first terms in its Taylor series expansion with that of the exact solution; see Ref. \cite{Shchesnovich24}. 


\subsection{Quantum state in the isoenergetic model}  

Let us now derive the state of the system for the coherent pump input with amplitude $\alpha$ by using the form of the amplitudes obtained in Eq. (\ref{Sol}). Extending the approximate solution from $N\in \Omega_\epsilon(\alpha)$ to all the subspaces $\mathcal{H}_N$ (such an extension is done at the cost of the error $\mathcal{O}(\epsilon)$) and in each subspace to $n\le N$, to obtain a simpler final expression, we derive from Eq. (\ref{Sol}) the following state
\begin{eqnarray}
\label{sol_coh}
 && e^{-\frac{it}{\hbar} {H}}|\alpha\rangle|0\rangle\approx e^{-\frac{\alpha^2}{2}}\sum_{N=0}^\infty  \frac{\alpha^{N}}{\sqrt{N!}} \nonumber\\
&& \times \sqrt{\sech r_N}\sum_{n=0}^{N}\binom{2n}{n}^\frac12  \left(\frac{  \tanh r_N}{2 }\right)^n |N-n,2n\rangle,\nonumber\\
 \end{eqnarray}
whose applicability domain is defined by the validity of the solution in Eq. (\ref{Sol}) discussed below. 
 
The state in Eq. (\ref{sol_coh}) describes the entanglement between the signal and pump modes. The entanglement results in a signal state with a fluctuating parameter $r_N^2$ about its average value $\overline{r_N^2} =r^2=(2\alpha \tau)^2$, where the overline denotes the average over the Poisson distribution $\mathcal{P}_N = e^{-\alpha^2}\alpha^{2N}/N!$. 

The above interpretation is corroborated by the fact that the average number of photons in the signal and the dispersion of the quadratures $X_+= (b+b^\dag)/2$ and $X_-=-i(b+b^\dag)/2$ are formally similar to their values within the standard parametric approximation, but with a fluctuating squeezing parameter $r_N$ [up to an error $\mathcal{O}(r/\alpha^2)$]:
\be
\langle b^\dag b\rangle = \overline{\sinh^2 r_N},\quad 
\langle \Delta X^2_\pm\rangle =  \overline{\frac{e^{\pm 2r_N}}{4}}.
\en{averADD}
Indeed, by the parity of the Fock states in the signal mode in Eq. (\ref{sol_coh}), we have exactly $\langle b\rangle = \langle b^\dag \rangle = 0$, so $\langle X_\pm \rangle = 0$. 
Averaging over the Poisson distribution in the confidence interval $N\in\Omega_\epsilon(\alpha)$ and using the fact that the quantum amplitudes in Eq. (\ref{Sol}) are formally similar to those of a Gaussian state with the squeezing parameter $r_N$, we obtain (see details in Appendix \ref{sec:AppendixIEA}):
\be 
\langle b^\dag b\rangle = \overline{\sinh^2 r_N}= \sinh^2 r+ \frac12\sum_{p=1}^\infty \frac{(2r)^{2p}}{(2p)!}h_p\left(\frac{r}{2\alpha}\right),
\en{avern}
with  the  coefficients  
\[
h_p(x) \equiv \sum_{i=1}^\infty {p+i\brace p}\frac{(4x)^{2i}}{(2p+1)\ldots (2p+2i)},
\]
where  ${k\brace p}$ is  the Stirling number of the second kind. An approximate expression for the mean number of signal photons
[up to an error $\mathcal{O}(r/\alpha^2)$] reads
\begin{eqnarray}
\label{avern_A}
 \langle b^\dag b\rangle =\overline{\sinh^2 r_N} \approx e^{\frac{r^2}{2\alpha^2}}\sinh^2r  + \frac12\left(e^{\frac{r^2}{2\alpha^2}}-1\right).
\end{eqnarray} 
Similarly  we get [up to an error  $\mathcal{O}(r/\alpha^2)$]:
\be
 \langle\Delta X^2_\pm\rangle = \overline{\frac{e^{\pm 2r_N}}{4}} \approx \frac14 e^{\pm 2r +\frac{r^2}{2\alpha^2}}. 
\en{aver2n} 

Observe that, independently of the pump state, the correction to the average number of signal photons given by the parametric approximation is always positive. This qualitative feature has a very simple explanation from the form of the solution in Eq. (\ref{Sol}), which  has the  amplitudes in subspace $\mathcal{H}_N$ that coincide with those of the standard Gaussian state with the squeezing parameter $r_N$. Thus, the whole solution in Eq. (\ref{sol_coh}) is a squeezed state with a fluctuating squeezing parameter about the mean value $r$. Since we average $\sinh^2r_N$ with positive Taylor coefficients in the expansion in powers of $r_N^2=4 N \tau^2$,  the correction to the number of photons in the signal mode is always positive. On the other hand, the correction to the variance of the squeezed qudrature in Eq.~(\ref{aver2n}) is negative, but leads to a saturation of squeezing at a much higher $r=2\alpha^2$ than the perturbative solution, far beyond the validity domain. We distinguish one more remarkable feature of the isoenergetic approximation: All odd moments of the quadratures are exactly zero, $\langle X^{2k+1}_\pm\rangle=0$, which is a consequence of the parity of the Fock states of the signal mode in Eq. (\ref{sol_coh}). 

We see that Eqs. (\ref{avern}) and (\ref{aver2n}) cannot reproduce the two main features of the perturbative solution \cite{Hillery84,Crouch88} discussed in Sec.~\ref{sec:Decomposition}. This fact shows   that the truncation of the perturbative series at the lowest-order correction term is dubious. At the same time, we should remember that the isoenergetic approximation cannot be extrapolated beyond its domain of validity, e.g., real saturation of squeezing can occur at a much lower $r$. Within the isoenergetic approach, we were able to solve the Schr\"odinger equation in the Fock space, in the limit of sufficiently strong pump power, only for the amplitudes $\Psi^{(N)}_n$   with $n\ll \sqrt{N}$, Eq. (\ref{Sol}). This solution becomes the exact solution in the formal limit of infinite power in the pump mode: The parameter $N$ in the $\beta$-coefficients in Eq. (\ref{Schr})  becomes formally independent of the size of the subspace (which was also $N$), whereas the subspace size is considered to be infinite (see discussion under Eq. (\ref{Eqfn})). Thus, our isoenergetic approximation is an improved version of the parametric approximation, which is consistent with the energy conservation. We will see below that indeed the former brings an error quadratic in respect to that of the latter.


\subsection{Validity domain of the isoenergetic approximation \label{sec:Validity}}

We have found the solution, Eq. (\ref{Sol}), to the Schr\"odinger equation which captures the first $n_N\approx \sqrt{\epsilon N}$ Fock state amplitudes in the subspace $\mathcal{H}_N$ to a relative error $\epsilon$. Using such an approximation in the subspaces $\mathcal{H}_N$ with $N\in \Omega_\epsilon(\alpha)$ results, to the same error $\epsilon$, in the state of Eq. (\ref{sol_coh}) for a strong pump in a coherent state $|\alpha\rangle$. Since the state in Eq. (\ref{sol_coh}) is valid up to a cut-off in each subspace $\mathcal{H}_N$, there must be a condition of validity of such an approximation. In the plane $(r, \alpha)$, the domain where the solution of Eq. (\ref{sol_coh})  approximates the exact state to a given error $\epsilon$ can be   found by considering how the norm of the respective quantum state deviates from $1$ and how the discarded higher-order terms  $n>n_N=\sqrt{\epsilon N}$ would affect the average number of photons in the signal mode. These observations lead to the following two estimates.  
 
{(i)}. We can estimate the domain of validity of the state in Eq. (\ref{sol_coh}) to a relative error $\mathcal{O}(\epsilon)$ by postulating that the norm of the state in Eq. (\ref{Sol}), cut to $n\le n_N=\sqrt{\epsilon N}$, has the same error $\mathcal{O}(\epsilon)$ in the confidence interval $\Omega_\epsilon(\alpha)$ in Eq. (\ref{bound}). First, let us estimate the contribution of the imposed cut-off to $n\le n_N$. To the relative error on the order $\mathcal{O}\left({1}/{n_N}\right)$, we get (see details in Appendix \ref{sec:AppendixIEA}) 
 \begin{eqnarray}
\label{norm}
  \delta ||\Psi^{(N)}||^2\equiv 1-\sum_{n=0}^{n_N}|\Psi^{(N)}_n|^2    = \frac{(\tanh r_N)^{2n_N} }{\sech{r_N}\sqrt{\pi n_N}}. 
\end{eqnarray}
In Eq. (\ref{norm}) we can substitute $N\to \alpha^2$ and $r_N\to r$, which for $N\in \Omega_\epsilon(\alpha)$ of Eq. (\ref{bound}) results in a relative error of at most $\mathcal{O}(\sqrt{\epsilon}+1/\alpha)$. Using the fact that the average probability of getting no photon counts from Eq. (\ref{Sol}) reads $p_0(r) = \sech r$ (to a relative error $\mathcal{O}(|r_N-r|) = \mathcal{O}(1/\alpha)$ in $\Omega_\epsilon(\alpha)$) and $n_N=\sqrt{\epsilon N}\approx \sqrt{\epsilon}\alpha$, we obtain that Eq. (\ref{sol_coh}) approximates the actual state with error $\epsilon$ in the probability distribution, i.e., the order of $\delta||\Psi^{(N)}||^2$ in  Eq. (\ref{norm}) is at most $\epsilon$, under the following condition 
 \be
V_1(r,\alpha,\epsilon)\equiv \frac{[1-p^2_0(r)]^{\sqrt{\epsilon} \alpha}}{p_0(r) \sqrt{\pi \alpha}  \epsilon^\frac{5}{4}}<1.
 \en{validity1}

{(ii)}. We can estimate the domain of validity of the state in Eq. (\ref{sol_coh}) to a relative error $\mathcal{O}(\epsilon)$ by postulating that the average photon number differs from that of the exact solution by the same relative error $\mathcal{O}( \epsilon)$.  Observing that the quantum amplitudes in Eq. (\ref{Sol}) for $0\le n\le n_N$ give 
\be
\langle\Psi^{(N)}|  {b}^\dag {b}|\Psi^{(N)}\rangle= \sinh^2r_N - \sum_{n=n_N+1}^\infty 2n|\Psi^{(N)}_n|^2,
\en{observ}
and estimating the summation term (see details in Appendix \ref{sec:AppendixIEA}), we obtain (to a relative error on the order $\mathcal{O}\left({1}/{n_N}\right)$) 
\begin{eqnarray}
\label{avPSI}
\sum_{n=0}^{n_N}|\Psi^{(N)}_n|^22n&=& \sinh^2r_N  - 2n_N\left[1-\sum_{n=0}^{n_N}|\Psi^{(N)}_n|^2  \right] \nonumber\\
& =&  \sinh^2r_N  - 2\sqrt{n_N}\frac{(\tanh r_N)^{2n_N} }{\sech{r_N}\sqrt{\pi }}.
 \end{eqnarray}

From Eq. (\ref{avPSI}) with the help of Eq. (\ref{norm}) we can obtain the condition for the discarded tail $n> n_N$  of the state in Eq. (\ref{Sol}) to contribute a relative error $\mathcal{O}(\epsilon)$ to the average number of photons in the signal mode. By substituting $N\to \alpha^2$ and $r_N\to r$ in Eq. (\ref{avPSI}), at the cost of a relative error at most $\mathcal{O}(\sqrt{\epsilon}+1/\alpha)$ (similar as in the norm estimate), we obtain the following condition   
\be
V_2(r,\alpha,\epsilon)\equiv \frac{2 \sqrt{\alpha}}{\sqrt{\pi} \epsilon^\frac34}p_0(r) [1-p^2_0(r)]^{\sqrt{\epsilon} \alpha-1} <1.
\en{validity2} 
Both conditions, Eqs. (\ref{validity1}) and (\ref{validity2}), are satisfied in the region 
\be
V_{IE}(r,\alpha,\epsilon)\equiv \max\{V_1,V_2\} <1.
\en{validity3} 
The derivations leading to Eq. (\ref{validity3}) implied that we discard the terms on the order of $\delta\equiv 1/n_N = 1/\sqrt{\epsilon N}\sim 1/(\sqrt{\epsilon}\alpha)$. Therefore,  very small (desired) errors, $\epsilon \lesssim 1/\alpha^2$, or relatively small pump powers, $\alpha^2 \lesssim 1/\epsilon$, violate the applicability of the validity domain following from Eq. (\ref{validity3}).

\begin{figure}[!ht]
\centering
\includegraphics[width=\linewidth]{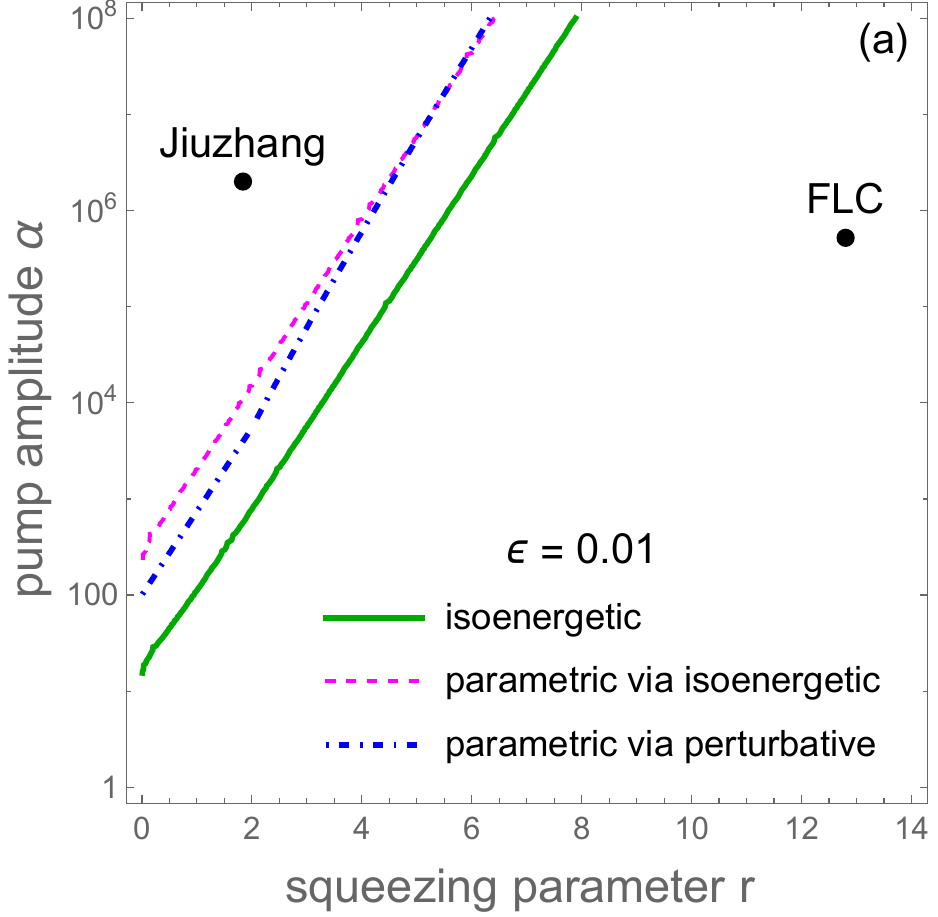}
\includegraphics[width=\linewidth]{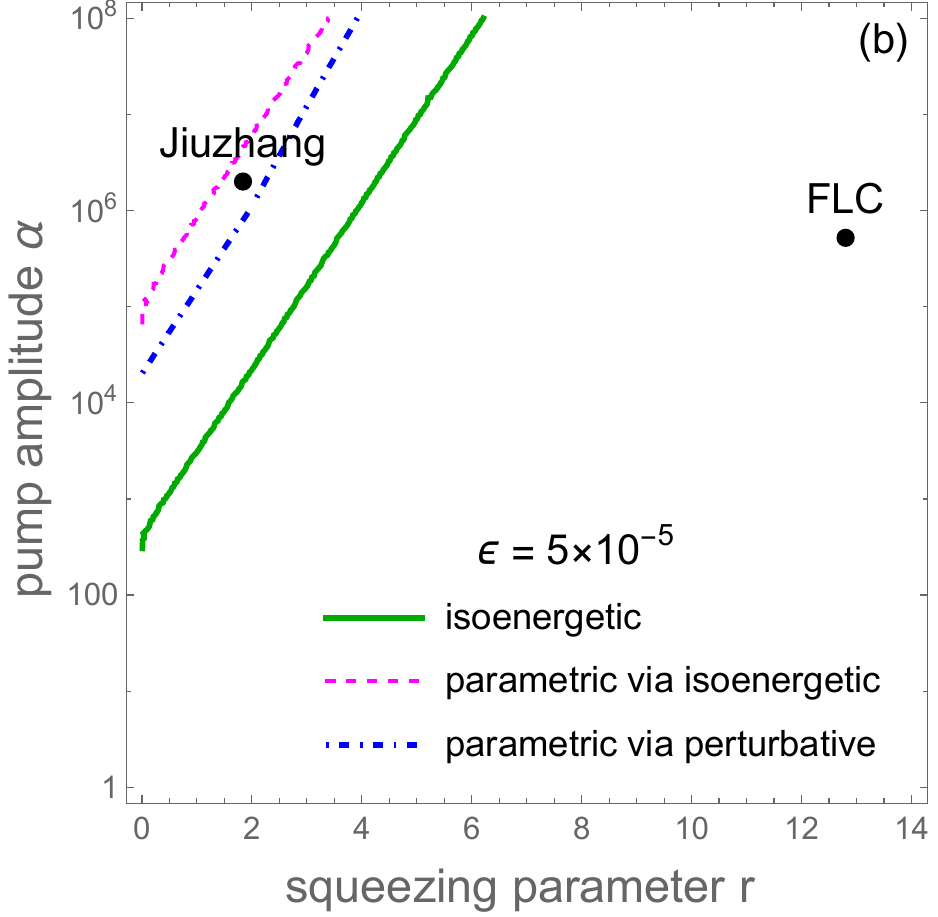}
\caption{Domains of validity of the parametric and isoenergetic approximations. The area above the green solid line corresponds to $V_{IE}(r,\alpha,\epsilon)<1$; in this area, the isoenergetic approximation is valid up to error $\epsilon$. The area above the magenta dashed line corresponds to $V_{P}(r,\alpha,\epsilon)<1$; in this area, the parametric approximation is valid up to error $\epsilon$ according to the isoenergetic approach. The area above the blue dot-dashed line corresponds to $V_{HZ}(r,\alpha,\epsilon)<1$; in this area, the parametric approximation is valid up to error $\epsilon$ according to the perturbative path integration of Hillery and Zubairy \cite{Hillery84}. 
The two points correspond to the experimental conditions of the Jiuzhang boson sampler \cite{Zhong20} and to the FLC experiment \cite{Florez20}, similar to Fig.~\ref{fig:ValidityPert}. \label{fig:ValidityIso}}
\end{figure}
 
Similar conditions as in Eqs. (\ref{validity1}) and (\ref{validity2}) can be found for the usual Gaussian state in the parametric approximation by a similar analysis of Eq. (\ref{PsiG}). By the above discussion, the domain of validity with the relative error $\epsilon$ given by Eqs. (\ref{validity1}) and (\ref{validity2}) correspond to the validity domains of the parametric approximation but with the relative error $\mathcal{O}(\sqrt{\epsilon})$. One can also see this in a straightforward way by direct reduction of the approximation of Eq. (\ref{sol_coh}) to the Gaussian squeezed state. Tracing out the pump in Eq. (\ref{sol_coh}), we obtain to an error $\epsilon$ the density matrix of the signal mode  
 \begin{eqnarray}
\label{rho}
&& {\rho}= \sum_{n=0}^{\infty} \sum_{m=0}^{\infty} 
\left[\binom{2n}{n}\binom{2m}{m}\right]^\frac12 \nonumber\\
&&\times \sum_{N=0}^{\infty}\mathcal{P}_N(\alpha)\alpha^{n+m} \left[(N+1)^{(n)}(N+1)^{(m)}\right]^{-\frac12}\nonumber\\
&&\times  \frac{\left(\frac{\tanh r_{N+n}}{2 } \right)^n \left(\frac{\tanh r_{N+m}}{2 } \right)^m}{\left[\cosh r_{N+n}\cosh r_{N+m}\right]^\frac12} |2n\rangle\langle 2m|,\nonumber\\
&& \end{eqnarray}
where  \mbox{$(N+1)^{(n)} \equiv   (N+1)\ldots (N+n)$}, etc. For $N\in \Omega_\epsilon(\alpha)$  similarly as in Eq. (\ref{err2}), we can use an asymptotic estimate as in Eq. (\ref{factor}),  with  $(N)_n$ replaced by $(N)^{(n)}$ and the minus sign removed from the exponent \cite{Shchesnovich21}, to obtain 
\be \frac{\alpha^n}{\sqrt{N^{(n)}}} =   1+ \mathcal{O}\left(\sqrt{\ln (1/\epsilon)}\frac{n}{\alpha} \right) = 1+ \mathcal{O}(\sqrt{\epsilon})
\en{est2}
and the previous estimate     $\tanh^N r_N- \tanh^nr = \mathcal{O}(\sqrt{\epsilon})$. Hence, the conditions for the usual Gaussian approximation to have an error $\epsilon$ are mapped to analogous conditions in Eq. (\ref{validity1})-(\ref{validity2}) for the solution in Eqs. (\ref{Sol}) and (\ref{rho}), but for the higher-order error $\epsilon^\prime= \epsilon^2$. Thus, the error-specific region of validity of the parametric approximation is
 \be
V_{P}(r,\alpha,\epsilon)\equiv V_{IE}(r,\alpha,\epsilon^2)<1.
 \en{validityP} 

The validity domains of the two approximations are compared in Fig. \ref{fig:ValidityIso} for various values of acceptable error $\epsilon$. We see in Fig.~\ref{fig:ValidityIso}(a) that the difference between the validity regions of the parametric approximation deduced from the isoenergetic approach and that obtained by Hillery and Zubairy from the perturbative expansion of path integration is small. However, at a lower error $\epsilon=5\times10^{-5}$, this difference becomes important, as we see in Fig.~\ref{fig:ValidityIso}(b). In particular, the Jiuzhang experiment is within the validity region after the traditional approach of Hillery and Zubairy, but it is outside the validity region determined by our refined method. 

\begin{figure}[!h]
\centering
\includegraphics[width=\linewidth]{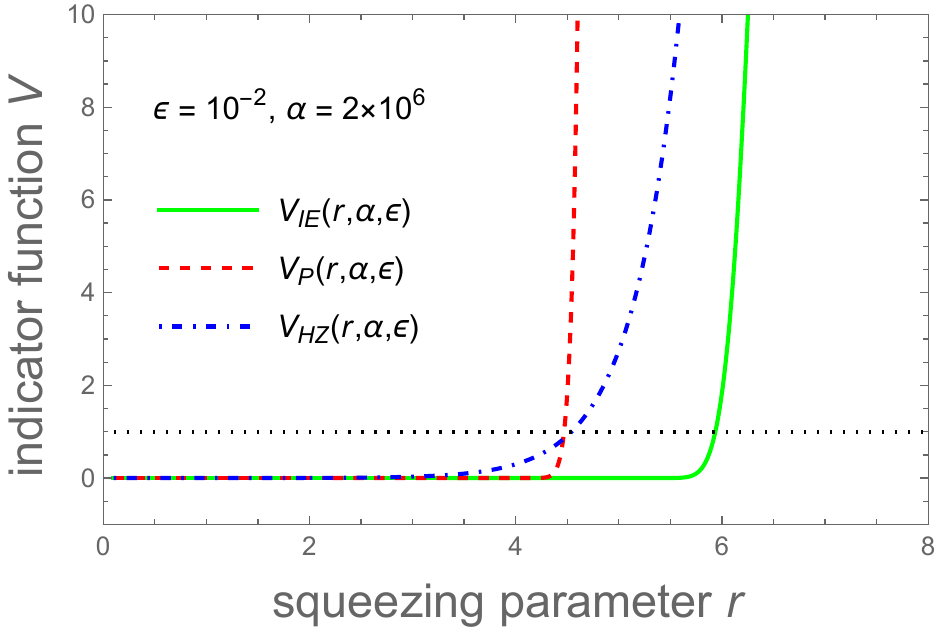}
\caption{Dependence of the indicator function on the squeezing parameter at a fixed pump amplitude. The indicator function for the isoenergetic approximation, $V_{IE}(r,\alpha,\epsilon)$, crosses the value 1 at a higher $r$ than both indicator functions for the parametric approximation. \label{ValidityV}}
\end{figure}

Figure~\ref{ValidityV} shows how fast the indicator function $V$ changes with $r$ at a fixed value of the pump amplitude, corresponding to the Jiuzhang boson sampler \cite{Zhong20}. This dependence can be obtained, for example, by changing the crystal length at a fixed pump power. We see that the parametric approximation is valid up to $r\approx4.5$, while the isoenergetic approximation is valid up to $r\approx6$. Above these values, the indicator function grows very fast, showing that the corresponding approximation becomes completely inapplicable.
 
\subsection{Numerical simulations}

In this section, we verify the validity of the isoenergetic approximation in the region defined by Eq. (\ref{validity3}) by direct numerical simulations of the full quantum model represented by
Eqs. (\ref{Ham}) and (\ref{Amp}). Numerical simulations can be performed independently in each invariant subspace. Since the only difference between them is the dimension parameter of the subspace $N$, performing numerical simulations in a single subspace that satisfies Eq. (\ref{bound}) is sufficient to draw conclusions. 

There is a vast literature on numerical simulations of dynamics for the closely related \emph{trilinear} Hamiltonian of a nondegenerate parametric amplifier, where the pump amplitude varies from $\alpha=7$ \cite{Walls70} and $\alpha=20$ \cite{Bandilla00} to $\alpha\approx300$ \cite{Chinni24}, that is, for much weaker pump powers than those in the current experiments with $\alpha \sim 10^6$. Higher values of $\alpha$ are not numerically achievable due to the growth of the computation complexity beyond the capabilities of modern computers. Similarly, our simulations are limited to relatively weak values of $\alpha$ due to the same limitation.    

We perform numerical simulations in the central invariant subspace $\mathcal{H}_N$, with $N = \alpha^2$ ($r_N = r$) for two values: $N = 4000$ ($\alpha \approx 63$) and $N=9000$ ($\alpha \approx 95$). We compute the Euclidean norm of the approximate state, 
\begin{equation}
||\Psi^{(N)}|| \equiv \langle \Psi^{(N)}|\Psi^{(N)}\rangle^{\frac12} =  \left(\sum\limits_{n=0}^{N} |\Psi^{(N)}_n|^2\right)^\frac12,   
\end{equation}
where we use only the leading-order term of $\Psi^{(N)}_n$  from Eq.~(\ref{Sol}) for the isoenergetic approximation and a similar norm with $\widetilde{\Psi}^{(N)}_n$ defined by Eq.~(\ref{PsiG}) for the parametric approximation. We also compute the numerical approximation error (the distance in norm) between the approximate state and the numerical solution $\Psi^{(N)}_\text{num}$ of the Schr\"odinger equation, Eq. (\ref{Schr}):
\begin{equation}
||\Delta\Psi^{(N)}|| = ||\Psi^{(N)}-\Psi^{(N)}_\text{num}||.   
\end{equation}

The results of the numerical simulations are presented in Tab.~\ref{tab:Norms} for the two values of $N$. Let us denote by $r_c(\epsilon,\alpha)$ the maximal value of the squeezing parameter where the isoenergetic approximation has an error below or equal to $\epsilon$. For sufficiently high values of $\alpha$, when the approach of Sec.~\ref{sec:Validity} is valid, $r_c$ is defined by the equation $V_{IE}(r_c,\alpha,\epsilon)=1$. At $\epsilon=0.01$, Eq. (\ref{validity3}) predicts $r_c(0.01,\sqrt{4000})\approx 0.7$, and $r_c(0.01,\sqrt{9000})\approx0.92$. The parametric approximation has the error $\sim \sqrt{\epsilon}=0.1$ for the same values of $r_c$. 

\begin{table*}
\caption{\label{tab:Norms} Hilbert space norm of the projection of the  quantum state onto $\mathcal{H}_N$ from two approximations, the isoenergetic and parametric, and the norm of the difference between the   respective  approximation and   the  numerical solution of the Schr\"odinger equation.}
\begin{ruledtabular}
\begin{tabular}{ccccccccc}
 &\multicolumn{4}{c}{$N=4000$}&\multicolumn{4}{c}{$N=9000$}\\
 &\multicolumn{2}{c}{Isoenergetic}&\multicolumn{2}{c}{Parametric}&\multicolumn{2}{c}{Isoenergetic}&\multicolumn{2}{c}{Parametric}\\
 $r$ &  $||\Delta \Psi^{(N)}|| $ &   $|| \Psi^{(N)}||$  &  $||\Delta \widetilde{\Psi}^{(N)}||$ &  $|| \widetilde{\Psi}^{(N)}||$ &  $||\Delta \Psi^{(N)}|| $ &   $|| \Psi^{(N)}||$  &  $||\Delta \widetilde{\Psi}^{(N)}||$ &  $|| \widetilde{\Psi}^{(N)}||$\\ \hline
 
 0.50 & 0.0001  &  1.00 &  0.0007 & 1.00 \\  
0.75 & 0.0004 & 1.00 &  0.007 & 1.00 \\ 
1.00 & 0.001 &  1.00 &  0.05 & 0.99 & 0.0006  &  1.00 &  0.023  & 0.99\\  
1.25 & 0.004  & 1.00 & 0.18 & 0.93 & 0.002 & 1.00 & 0.11 & 0.96\\ 
1.50 & 0.01 &  1.00 & 0.42  & 0.82 & 0.005 &  1.00 & 0.32  & 0.87\\ 
1.75 & 0.05 &  0.98 & 0.63  & 0.68 & 0.063 &  1.00 & 0.55  & 0.74\\ 
2.00 & 0.26 &  0.95 & 0.78 & 0.54 & 0.28 &  0.95 & 0.73 & 0.59\\
\end{tabular}
\end{ruledtabular}
\end{table*}

We see from the values of $||\Delta \Psi^{(N)}||$ in Tab.~\ref{tab:Norms}, that the error exceeds the acceptable error level of 0.01 at $r_c\ge1.5$ for both used values of $N$. That is, the numerical simulation gives the critical values $r_c$, which are higher than the predicted values (i.e., the applicability domain is wider than predicted). These values of $r_c$ can be trusted because the norm of the isoenergetic approximation, $|| \Psi^{(N)}||$, is still unit at these values of the squeezing parameter.
To understand this difference we need to remember that the criterion in Eq. (\ref{validity3}) was derived by employing an asymptotic approximation, which improves as $\alpha$ grows. 
The criterion does not account for some higher-order terms, starting from the order $\delta= 1/n_N$, where $n_N= \sqrt{\epsilon N}$, discarded in the derivations in Appendix~\ref{sec:AppendixIEA}; for $N$ in the numerical simulations and $\epsilon=0.01$ we obtain:  $\delta \approx  0.16$ for $N=4000$ and  $\delta\approx  0.1$ for $N=9000$. These values of $\delta$ exceed $\epsilon$ and therefore make imprecise the asymptotic formula Eq. (\ref{validity3}).

From the values of $||\Delta \widetilde{\Psi}^{(N)}||$ in Tab.~\ref{tab:Norms}, we see that the error of the parametric approximation is approximately square root of the error of the isoenergetic approximation when the norm is unit, and even higher when the norm is degrading. Thus, the numerical results confirm the error relation of the two approximations established in Sec.~\ref{sec:Validity}.

\subsection{Practical value}
When a single-mode squeezer is used for the generation of squeezed light, it is generally assumed that the parametric approximation is valid under the condition that the mean number of signal photons is much less than the mean number of pump photons, i.e., while the pump depletion is negligible. We have shown that this assumption is justified only in the simplest experimental conditions, for example, when just one squeezer is used and the quantity of interest is the average number of signal photons. However, in a more complicated setup, where the fields generated by multiple squeezers are mixed in a large-size interferometer with a subsequent measurement of optical modes at its output, as in continuous-variable optical quantum computers \cite{Larsen19,Asavanant19} and Gaussian boson samplers \cite{Aaronson11,Lund14,Hamilton17}, the restrictions for the applicability of the parametric approximation are much more stringent.  

To verify the applicability of the parametric approximation to the output state of a squeezer, one should measure the average energies of the pump pulse $U_p$ and the signal pulse
$U_s$, find the values of $\alpha$ and $r$ by solving equations $\alpha^2=U_p/\hbar\omega_p$ and $\sinh^2r=U_s/\hbar\omega_s$, and then calculate the value $V_P(r,\alpha,\epsilon)$ for the acceptable error level $\epsilon$, determined, for example, by the error-correction codes applied. If the value obtained is less than 1, the parametric approximation is applicable to this squeezer for whatever large number of interacting modes in the setup. If $V_P(r,\alpha,\epsilon)$ is not less than 1, but $V_{IE}(r,\alpha,\epsilon)$ is, then the parametric approximation can fail for this level of error, and the isoenergetic approximation should be used instead for the description of the output state of the entangled pump and signal beams of this squeezer.

\section{Conclusion}

We have analyzed the joint quantum state of the pump and signal modes in the down-conversion process within the strong pump regime, using both a perturbative approach and an approximate solution to the Schrödinger equation. By enforcing exact conservation of the optical energy in each invariant subspace of the Hilbert space, we have established, for the first time, precise conditions on the model parameters under which the standard parametric approximation remains valid. Furthermore, by solving the Schrödinger equation in the strong pump approximation while maintaining optical energy conservation, we have introduced a new isoenergetic approximation, which describes the signal mode as a quantum state with a fluctuating squeezing parameter. This approximation has the validity domain much wider than that of the parametric approximation and accounts for pump depletion and entanglement between the pump and signal modes.

Our perturbative approach has produced the corrections to the parametric approximation previously derived by Hillery and Zubairy using a path-integral approach \cite{Hillery84}. However, the corrections obtained from our approximate Schrödinger equation solution, which enforces optical energy conservation, do not match those from the perturbative approach. This discrepancy suggests potential limitations in either the isoenergetic approach — indicating the need for higher-order refinements — or the perturbative approach, whose validity domain cannot be determined from within the approach itself as the convergence of the perturbative series cannot be proven for unbounded operators.

We have also derived the validity domains for both parametric and isoenergetic approximations on the parameter plane $(r, \alpha)$. The parametric approximation remains valid in certain recent experiments, including Gaussian boson sampling \cite{Zhong20, Zhong21, Deng23}, where we confirm that the quantum state can still be well approximated as Gaussian, with an error below a fraction of a percent -- sufficiently small to meet the stringent error sensitivity requirements of Gaussian boson sampling protocols \cite{Qi20, Shchesnovich22}.

However, for future experiments with stronger squeezing, we predict a breakdown of the parametric (Gaussian-state) approximation. Specifically, for pump powers around $\alpha \approx 2\cdot10^6$, the Gaussian approximation is expected to fail for squeezing parameters $r \geq 4.5$. Although the isoenergetic approximation provides a more robust description, it is also expected to fail at higher squeezing values, around $r \geq 6$ for the same pump power. Moreover, recent progress in dispersion-engineered nanophotonics \cite{Jankowski21} allows one to hope for a significant reduction of pump power in squeezed state generation, which will result in a lower limit of validity of the parametric approximation, at the level of values practically used of $r\approx2$. 

To accurately describe the system outside the validity domain of the isoenergetic approximation, more precise approximations must be developed within the framework proposed here. Alternatively, one could profit from the exact solution available in Ref. \cite{Shchesnovich24}.


\section*{Acknowledgments}
The authors are grateful to Mikhail Kolobov and Nicolas Treps for illuminating discussions. V.S. thanks the National Council for Scientific and Technological Development of Brazil (CNPq) for financial support. D.H. acknowledges support from QuantERA II Programme (project EXTRASENS) that received funding from Horizon 2020 Framework Programme of the European Union, Grant No. 101017733 and the Bundesministerium für Bildung und Forschung (Germany), Grant No. 13N16935.

\appendix
   
\section{The norm of the projection of the  parametric approximation }
 \label{sec:AppendixPA}

By Eq. (\ref{PsiG}) we have to estimate for $\alpha\gg 1$ the following sum 
\be
\langle \widetilde{\Psi}^{(N)}|\widetilde{\Psi}^{(N)}\rangle = \sech r \sum_{n=0}^N \binom{N}{n}\frac{(2n)!}{n!}\left(\frac{\tanh r}{2\alpha}\right)^{2n}.
\en{eB1}
To this goal we use the identity
\be
\int\limits^\infty_{-\infty}\frac{\rd x }{\sqrt{\pi}}e^{-x^2}x^{2n} = \frac{(2n)!}{4^nn!}. 
\en{eB2}
Thus 
\begin{eqnarray}
\label{eB3}
&& \langle\widetilde{\Psi}^{(N)}\widetilde{\Psi}^{(N)}\rangle =\sech r \int\limits_{-\infty}^\infty \frac{\rd x }{\sqrt{\pi}}e^{-x^2} 
\sum_{n=0}^N \binom{N}{n} \left(\frac{x\tanh r}{\alpha}\right)^{2n}\nonumber\\
&&=\sech r \int\limits_{-\infty}^\infty \frac{\rd x }{\sqrt{\pi}}e^{-x^2} 
\left[ 1 +  \left(\frac{x\tanh r}{\alpha}\right)^{2}\right]^N. 
\end{eqnarray}
Setting $Z \equiv x\tanh r/\alpha$ and observing that for   strong pump powers     $|Z|\ll 1$, we approximate the second factor in the integrand as follows   
\begin{eqnarray}
\label{eB4}
&&\left( 1 + Z^{2}\right)^N = \exp\left\{N \ln \left( 1 + Z^2 \right)\right\}\nonumber\\
&&=\exp\left\{N \left[Z^2 +\mathcal{O}\left(Z^4 \right)\right] \right\}.
\end{eqnarray}
Retaining only the quadratic term in the exponent in Eq. (\ref{eB4}) and integrating, we obtain
\begin{eqnarray}
\label{eB5}
&&\langle\widetilde{\Psi}^{(N)}\widetilde{\Psi}^{(N)}\rangle \approx \frac{\sech r}{\sqrt{1 - \tanh^2 r N/\alpha^2 }}\nonumber\\
&& = \frac{1}{1 - \sinh^2 r \left(\frac{N}{\alpha^2}-1\right)}\approx  1 + \frac{\sinh^2 r}{2} \left(\frac{N}{\alpha^2}-1\right).\nonumber\\
\end{eqnarray}

The average of the norm in the invariant subspaces is exactly equal to $1$, as indicated in the main text. To obtain an estimate of the average error to the norm, we will average the squared difference. Introducing rescaled integration variables $x_i = \xi_i\cosh r $ in each factor, we get from Eq. (\ref{eB3}) the double integral representation  
\begin{eqnarray}
\label{EB4}
&& \left(1-\langle\widetilde{\Psi}^{(N)}\widetilde{\Psi}^{(N)}\rangle\right)^2 =\int\limits_{-\infty}^\infty \int\limits_{-\infty}^\infty \frac{\rd \xi_1\rd \xi_2 }{\pi}e^{-(\xi_1^2+\xi_2^2)\cosh^2r} 
\nonumber\\
&&\times\left[\cosh r - \left( 1 +  \left(\frac{\xi_1\sinh r}{\alpha}\right)^{2}\right)^N\right]\nonumber\\
&&\times \left[\cosh r -  \left( 1 +  \left(\frac{\xi_2\sinh r}{\alpha}\right)^{2}\right)^N  \right].
\end{eqnarray}
Averaging over the Poisson distribution of Eq. (\ref{CohSt})  in Eq. (\ref{EB4}) with the help of the identity  $\overline{X^N}=\exp\{-\alpha^2(1-X)\}$ (for an $N$-independent  $X$),  we obtain
\begin{eqnarray}
\label{EB5}
&& \overline{\left(1-\langle\widetilde{\Psi}^{(N)}\widetilde{\Psi}^{(N)}\rangle\right)^2} =\int\limits_{-\infty}^\infty \int\limits_{-\infty}^\infty \frac{\rd \xi_1\rd \xi_2 }{\pi}e^{-(\xi_1^2+\xi_2^2)\cosh^2r} 
\nonumber\\
&&\times\biggl( \cosh^2  r - \cosh r \left[ e^{\xi_1^2 \sinh^2 r } +   e^{\xi_2^2 \sinh^2 r }\right] ]\nonumber\\
&&+  \exp\left\{(\xi_1^2 +\xi_2^2) \sinh^2 r + \frac{\xi_1^2\xi_2^2 \sinh ^4 r}{\alpha^2}  \right\}  \biggr).
\end{eqnarray}
The first three terms in Eq. (\ref{EB5})  are Gaussian integrals, which can be integrated exactly, whereas the last exponent can be approximated for a strong pump by expanding over $1/\alpha^2$. Retaining the first-order term in the expansion, we then integrate in the polar coordinates [in the plane $(\xi_1,\xi_2)$] and  obtain
\begin{eqnarray}
\label{rBreak}
&&  \overline{\left(1-\langle\widetilde{\Psi}^{(N)}\widetilde{\Psi}^{(N)}\rangle\right)^2} \approx \frac{\sinh^4 r}{4\alpha^2}. 
\end{eqnarray}
Equation (\ref{rBreak}) is derived by assuming that the standard parametric approximation applies. For instance, it is assumed that Eq. (\ref{PsiG})  accounts for the exact amplitudes $\Psi^{(N)}_n$ for large $n$ in the interval $0\le n\le N$, which is not the correct assumption, as shown in the main text.   It is interesting that the order of the correction in Eq. (\ref{rBreak}) is the same as that of Ref.  \cite{Hillery84}.

\section{The norm of the approximation and the average number of photons}
\label{sec:AppendixIEA}
 
We will need the Stirling-type bound for the factorial $m! = \left(\frac{m}{e}\right)^m \sqrt{2\pi(m+\mu_m)}$,  where $1/6< \mu_m <1/3 $  \cite{Mortici11}. Applied to the binomial it gives
\be
\binom{2n}{n} = \frac{4^n}{\sqrt{\pi n}}\left[1+\mathcal{O}\left(\frac{1}{n}\right)\right]. 
 \en{EB0}
 
Below we frequently use the fact that the amplitudes in Eq. (\ref{Sol}), if extended to infinite values of $n$, coincide with those of the standard Gaussian state with the squeezing parameter $r_N=2\sqrt{N}\tau$. They give exact quantum amplitudes in the subspace $\mathcal{H}_N$ only for $n\le n_N = \sqrt{\epsilon N}$, up to a relative error $\mathcal{O}(\epsilon)$.  Precisely this feature allows us to estimate the relative error of the isoenergetic approximation. 

We will give two estimates on the relative error: (i) we estimate the variation of the norm of the quantum state $|\Psi^{(N)}\rangle$  with the amplitudes from Eq. (\ref{Sol}) caused by discarding the photon numbers $n\ge n_N$ and (ii) we estimate the variation of the average number of photons in the signal mode caused by discarding the same photon numbers. Using the estimate in Eq. (\ref{EB0}) and the identity
 \be
 \int\limits_{-\infty}^{\infty}\frac{\rd z}{\pi}  e^{-nz^2} = \frac{1}{\sqrt{\pi n}},
 \en{EqB1}
to convert a sum to an integral, we obtain to a relative error on the order $\mathcal{O}\left({1}/{n_N}\right)$ :
\begin{eqnarray}
\label{normB}
&&1-\langle \Psi^{(N)}|\Psi^{(N)}\rangle  = \sech r_N \sum_{n=n_N}^{\infty} \frac{(\tanh r_N)^{2n}}{\sqrt{\pi n }} 
 \nonumber\\
 && = \sech r_N\int\limits_{-\infty}^{\infty}\frac{\rd z}{\pi} \sum_{n=n_N}^\infty e^{-nz^2}(\tanh r_N)^{2n} \nonumber\\
 && = \sech r_N \int\limits_{-\infty}^{\infty}\frac{\rd z}{\pi} \frac{\left(e^{-z^2}\tanh^2r_N \right)^{n_N}}{1-e^{-z^2}\tanh^2 r_N }  \nonumber\\
&&= \cosh r_N  \frac{(\tanh r_N)^{2n_N} }{\sqrt{\pi n_N}},
\end{eqnarray}
where the integral is approximated by changing the variable $\xi = \sqrt{n_N} z$  and expanding over $1/n_N$. 

Consider now the average number of photons in the state $|\Psi^{(N)}\rangle$.  Denoting $T \equiv \tanh r_N$, observing that the quantum amplitudes in Eq. (\ref{Sol}) give the average number of photons equal to $\sinh^2 r_N$, if formally extended to infinity, and using Eqs. (\ref{EB0})-(\ref{normB}), we obtain  to an error on the order $\mathcal{O}\left({1}/{n_N}\right)$:
\begin{eqnarray}
&&\sinh^2 r_N -  \langle\Psi^{(N)}| {b}^\dag {b}|\Psi^{(N)}\rangle =  \sech r_N   \sum_{n=n_N}^{\infty}\binom{2n}{n} \frac{T^{2n}}{4^n} 2n\nonumber\\
&& =   \sech r_N T\frac{\partial }{\partial T} \sum_{n=n_N}^{\infty}\binom{2n}{n} \frac{T^{2n}}{4^n} \nonumber\\
&& =    \sech r_NT\frac{\partial }{\partial T} \int\limits_{-\infty}^{\infty}\frac{\rd z}{\pi} \frac{\left(e^{-z^2} T^2 \right)^{n_N}}{1-e^{-z^2}T^2  }  \nonumber\\
&&=\sech r_N 2n_N   \int\limits_{-\infty}^{\infty}\frac{\rd z}{\pi} \frac{\left(e^{-z^2} T^2 \right)^{n_N}}{1-e^{-z^2}T^2  } 
\nonumber\\
&& =  2n_N \left( 1- \langle\Psi^{(N)} | \Psi^{(N)} \rangle\right). 
\label{EqB2}\end{eqnarray}
Equations (\ref{normB}) and (\ref{EqB2}) are used to estimate the validity domain of the isoenergetic approach in the main text. These results have been derived at the cost of discarding terms on the order of $\delta\equiv 1/n_N = 1/\sqrt{\epsilon N}\sim 1/(\sqrt{\epsilon}\alpha)$. 

Now consider the average number of photons in the state of Eq.~(\ref{sol_coh}) given by the averaging  over the Poisson distribution  
\mbox{$\mathcal{P}_N = e^{-\alpha^2}\alpha^{2N}/N!$} of the quantity   $ \langle \Psi^{(N)}|b^\dag b |\Psi^{(N)}\rangle$ with the state in Eq. (\ref{Sol}). Using the Taylor series expansion of $\sinh^2(x)$ 
and that the $k$th  moment of the Poisson distribution reads
\be 
\overline{N^k} = \sum_{p=0}^{k}{k\brace p}\alpha^{2p},
\en{Moment_k}
with 
\be 
{k\brace p}\equiv \sum_{l=0}^p \frac{(-1)^{p-l} l^{k}}{(p-l)! l!}
\en{Stirl2} 
being the Stirling number of the second kind,   we obtain  
\begin{eqnarray}
&& \langle b^\dag b\rangle = \overline{\sinh^2r_N}= \frac12\sum_{k=1}^\infty \frac{\overline{(2r_N)^{2k}}}{(2k)!} \nonumber\\
&& =\frac12\sum_{k=1}^\infty \frac{(4\tau)^{2k}}{(2k)!}\sum_{p=1}^{k}{k\brace p}\alpha^{2p}  = \sinh^2r+\frac12\sum_{p=1}^\infty \frac{(2r)^{2p}}{(2p)!}h_p(\tau),\nonumber\\ 
\label{EB8}\end{eqnarray}
where we have exchanged the order of summations (setting $k=p+i$) and subtracted the term with $i=0$ from the inner sum 
(using that ${k\brace 0}=\delta_{k,0}$ and  ${k\brace k}={k\brace 1}=1$ for all $k\ge 1$), and denoted  
\be
h_p(\tau) \equiv \sum_{i=1}^\infty {p+i\brace p}\frac{(4\tau)^{2i}}{(2p+1)\ldots (2p+2i)}.
\en{G_factor}

Similarly, we can evaluate the dispersion of the quadratures of the signal mode. Due to the definite parity of the Fock states (photon pairs) of the signal mode in the combined state, Eq. (\ref{sol_coh}), we have $\langle b\rangle =0$.   Since the quantum amplitudes are real, we also have $\langle b^\dag{}^2\rangle  = \langle b^2\rangle$. We have the following formula for the quadratures $X_+= \frac12(b+b^\dag)$ and $X_-=\frac{1}{2i}(b-b^\dag)$ 
\be 
\langle X^2_\sigma\rangle = \frac12\left(\langle b^\dag b\rangle + \sigma \langle b^2\rangle +\frac12\right),\quad  \sigma=\pm.  
\en{EB9}
Now consider the average on the state of Eq. (\ref{sol_coh}) 
\begin{eqnarray}
&&\langle b^2\rangle = \sum_{N=1}^\infty\sqrt{\mathcal{P}_N\mathcal{P}_{N-1}}\sum_{n=1}^N\Psi^{(N)}_n\Psi^{(N-1)}_{n-1}\sqrt{2n(2n-1)}\nonumber\\
&&=\sum_{N=1}^\infty\sqrt{\mathcal{P}_N\mathcal{P}_{N-1}} \tanh r_N \sum_{n=0}^{N-1}\Psi^{(N)}_n\Psi^{(N-1)}_{n} (2n+1),\nonumber\\
\label{EB10}\end{eqnarray}
where we have used that
\be
\Psi^{(N)}_n \sqrt{2n(2n-1)} = \tanh r_N \Psi^{(N)}_{n-1} (2n-1) 
\en{EB11}
and changed the index of summation $n\to n-1$. In Eq. (\ref{EB10}) we can neglect the difference between $N$ and $N-1$ in the sum over the Poisson distribution $\mathcal{P}_N = e^{-\alpha^2}\alpha^{2N}/N!$, since $\sqrt{\mathcal{P}_{N-1}} =\frac{\sqrt{N}}{\alpha}\sqrt{\mathcal{P}_N} = \left[1+\mathcal{O}\left( \frac{1}{\alpha}\right)\right]\sqrt{\mathcal{P}_N}$, in the interval $N\in \Omega_\epsilon(\alpha)$ and set $r_{N-1} = r_N$ in the resulting expression (for the same reason). Thus we obtain (up to a relative error $\mathcal{O}\left(\frac{r }{\alpha^2}\right)$)
\begin{eqnarray}
&&\langle b^2\rangle \approx  \sum_{N=1}^\infty \mathcal{P}_N  \tanh r_N \sum_{n=0}^{N-1}(\Psi^{(N)}_n)^2  (2n+1),\nonumber\\
&& = \overline{\tanh r_N( \sinh^2 r_N +1)}.
\label{EB12}\end{eqnarray}
From Eqs. (\ref{EB8})-(\ref{EB12}),  using trigonometric identity 
\be
\sinh^2x +\sigma \tanh x(\sinh^2 x+1) +\frac12 = \frac{e^{2\sigma x}}{2}, 
\en{EB13}
we arrive at the following result 
\begin{eqnarray}
\langle X^2_\sigma\rangle = \overline{\frac{e^{2\sigma r_N}}{4}}.
\label{EB14}\end{eqnarray}
To perform averaging, we have to have powers of $N$ and not of $\sqrt{N}$. For $N\in \Omega_\epsilon(\alpha)$, we can expand 
\begin{eqnarray}
\label{rN_approx}
 &&r_N = 2\sqrt{N}\tau =  r\sqrt{1+\frac{N}{\alpha^2}-1}  \nonumber\\
 &&=  \frac{r}{2}\left(1+ \frac{N}{\alpha^2}\right) + \mathcal{O}\left(\frac{r }{\alpha^2}\right),
 \end{eqnarray}
where we have used that $\mathcal{O}\left(\frac{r\ln(2/\epsilon)}{\alpha^2}\right) = \mathcal{O}\left(\frac{r }{\alpha^2}\right)$. With  the help of the identity  $\overline{Z^N}=\exp\{-\alpha^2(1-Z)\}$, when averaging over the Poisson distribution for an $N$-independent $Z$, we obtain [up to a relative error $\mathcal{O}\left(\frac{r }{\alpha^2}\right)$]
\begin{eqnarray}
&&\langle X^2_\sigma\rangle = \overline{\frac{e^{2\sigma r_N}}{4}}\approx \frac{e^{\sigma r}}{4}\exp\left\{-\alpha^2\left[1-e^{-\frac{\sigma r}{\alpha^2}}\right]\right\}\nonumber\\
&& \approx \frac14 e^{2\sigma r +\frac{r^2}{2\alpha^2}}.
\label{EB14A}\end{eqnarray}

Similarly, one can get a simple approximate expression for the average number of signal photons [up to a relative error $\mathcal{O}\left(\frac{r }{\alpha^2}\right)$]. Using the above mentioned identity $\overline{Z^N}=\exp\{-\alpha^2(1-Z)\}$ and Eq. (\ref{rN_approx}) we obtain:
\begin{eqnarray}
&& \langle b^\dag b\rangle = \overline{\sinh^2r_N}\approx   \overline{\sinh^2\left( \frac{r}{2} +\frac{rN}{2\alpha^2}\right)}\nonumber\\
&&=\frac14\left( \sum_{\sigma = \pm} e^{\sigma r}\exp\left\{-\alpha^2\left[1-e^{-\frac{\sigma r}{\alpha^2}}\right]\right\}\right)-\frac12 \nonumber\\
&& \approx e^{\frac{r^2}{2\alpha^2}}\sinh^2r  - \frac12\left(1-e^{\frac{r^2}{2\alpha^2}}\right). 
\label{EB8A}\end{eqnarray}

We also show that the kurtosis of the quadratures is exactly zero. For example, for $X^3_-$, we obtain  
\be
X^3_-= \frac{i}{8}\left\{3(b^\dag{}^2 b - b^\dag b^2 + b^\dag -b) + b^3 - b^\dag{}^3\right\},
\en{Kurt}
and, by the parity of the Fock states of the signal mode in Eq. (\ref{sol_coh}), we get $\langle X_-^3\rangle =0$. Similarly, $\langle X_+^3\rangle =0$.

\bibliography{Bib-Nonparametric}

\end{document}